\documentclass[preprint,12pt,sort&compress]{elsarticle}

\usepackage{amsmath,amssymb,mathrsfs}
\usepackage[]{graphicx}
\usepackage[latin1]{inputenc}
\usepackage{color}
\usepackage{enumerate}
\usepackage{hyperref}
\usepackage{caption}
\hypersetup{colorlinks=true,citecolor=blue}

\begin{document}

\begin{frontmatter}

\title{Monitoring crystal breakage in wet milling processes using inline imaging and chord length distribution measurements}
\author[a1]{Okpeafoh S. Agimelen\corref{cor1}}
\ead{okpeafoh.agimelen@strath.ac.uk}

\author[a1]{Vaclav Svoboda}

\author[a2]{Bilal Ahmed}

\author[a3]{Javier Cardona}

\author[a4]{Jerzy Dziewierz}

\author[a2]{Cameron J. Brown}

\author[a2]{Thomas McGlone}

\author[a3]{Alison Cleary}

\author[a3]{Christos Tachtatzis}

\author[a3]{Craig Michie}

\author[a2]{Alastair J. Florence}

\author[a3]{Ivan Andonovic}

\author[a5]{Anthony J. Mulholland}

\author[a1]{Jan Sefcik\corref{cor1}}
\ead{jan.sefcik@strath.ac.uk}

\cortext[cor1]{Corresponding authors}
\address[a1]{EPSRC Centre for Innovative Manufacturing in Continuous Manufacturing and Crystallisation, Department of Chemical and Process Engineering, University of Strathclyde, James Weir Building, 75 Montrose Street, Glasgow, G1 1XJ, United Kingdom.}

\address[a2]{EPSRC Centre for Innovative Manufacturing in Continuous Manufacturing and Crystallisation, Strathclyde Institute of Pharmacy and Biomedical Sciences, University of Strathclyde, 161 Cathedral Street, Glasgow, G4 0RE, United Kingdom.}

\address[a3]{Centre for Intelligent Dynamic Communications, Department of Electronic and Electrical Engineering, University Of Strathclyde, Royal College Building, 204 George Street, Glasgow, G1 1XW, 
United Kingdom.
}

\address[a4]{The Centre for Ultrasonic Engineering, Department of Electronic and Electrical Engineering, University Of Strathclyde, Royal College Building, 204 George Street, Glasgow, G1 1XW, 
United Kingdom.
}

\address[a5]{Department of Mathematics and Statistics, University of Strathclyde, Livingstone Tower, 26 Richmond Street, Glasgow, G1 1XH, United Kingdom.}

\begin{abstract}

The success of the various secondary operations involved in the production of particulate products depends on the production of particles with a desired size and shape from a previous primary operation such as crystallisation. This is because these properties of size and shape affect the behaviour of the particles in the secondary processes. The size and the shape of the particles are very sensitive to the conditions of the crystallisation processes, and so control of these processes is essential. This control requires the development of software tools that can effectively and efficiently process the sensor data captured in situ. However, these tools have various strengths and limitations depending on the process conditions and the nature of the particles.

In this work, we employ wet milling of crystalline particles as a case study of a process which produces effects typical to crystallisation processes. We study some of the strengths and limitations of our previously introduced tools for estimating the particle size distribution (PSD) and the aspect ratio from chord length distribution (CLD) and imaging data. We find situations where the CLD tool works better than the imaging tool and vice versa. However, in general both tools complement each other, and can therefore be employed in a suitable multi-objective optimisation approach to estimate PSD and aspect ratio.

\end{abstract}


\end{frontmatter}

\section{Introduction}
\label{sec1}

The success of any manufacturing process for particulate products depends on some key attributes of the particles which influence the outcome of various downstream operations carried out in the manufacturing process. In particular, the particle attributes of size and shape influence their behaviours such as flowability, filterability, ease of dissolution and so on. These behaviours in turn determine if the various downstream operations will be successful or not. Hence it is necessary that the particles possess the desired particle size distribution (PSD) and shape for a particular process \cite{Barrett2005,Paul2005,Chen2011}.

A very important upstream operation which is crucial to the manufacture of particulate products is the crystallisation process. The PSD and shape of the particles produced during this crystallisation process
vary due to a number of factors which include the nature of the material of which the particles are composed and the crystallisation process conditions. The combination of factors that will lead to the production of particles with the desired PSD and shape are not easily determined. Therefore, for the production of particles with the desired PSD and shape during crystallisation processes, it is necessary that the crystallisation process be controlled so as to produce tailor made particles. 

In order to control crystallisation processes, it is necessary that the PSD and shape of the particles produced in the process be monitored in situ, and this can be achieved by the use of inline sensors. However, these inline sensors have limitations on their applicability which could be due to the working principles of the sensors or various effects from the process or both. These limitations will influence the level of accuracy of the estimated PSD and aspect ratio (the metric for quantifying particle shape) obtained using these sensors. This will then determine if the estimated PSD and aspect ratio obtained from the sensor data is representative of the particles being measured.

In this work, we examine various process conditions and their effects on two inline sensors; the Mettler Toledo focused beam reflectance measurement (FBRM) and the particle vision and measurement (PVM) sensors. These two sensors have been chosen because of their wide applicability in various particulate processes. We employ these sensors in different wet milling processes of three different organic crystalline particles. Wet milling is one of the key processes carried out in industry during the production of particulate products. This process produces different effects typical of crystallisation processes. We explore the performance of these sensors under these effects.

\section{Methods}
\label{sec2}

The methodology employed in this work consisted of wet milling experiments and subsequent analysis of data acquired inline and offline. The materials, equipment and experimental procedure employed are described in subsections \ref{subsec2-1} to \ref{subsec2-3}, while the methods of data analysis used are described in subsection \ref{subsec2-4}.

\subsection{Materials}
\label{subsec2-1}

The following materials were used in this work: paracetamol (98.0-102.0\% USP), benzoic acid ($>99.5$\%), and metformin hydrochloride (reagent grade). Both paracetamol and benzoic acid were purchased from Sigma-Aldrich while metformin hydrochloride was purchased from Molekula. The benzoic acid particles were suspended in distilled water obtained from an in-house purification system, and the surfactant Tween 20 from Sigma-Aldrich was added to the benzoic acid slurry to ease dispersion of the particles and avoid foaming. However, the paracetamol and metformin hydrochloride were both suspended in 2-propanol (reagent grade, CAS: 67-63-0, Assay (GLC) $>99.5$\%) obtained from from Fisher Scientific, UK.

\subsection{Equipment}
\label{subsec2-2}

\begin{figure}[tbh]
\centerline{\includegraphics[width=0.5\textwidth]{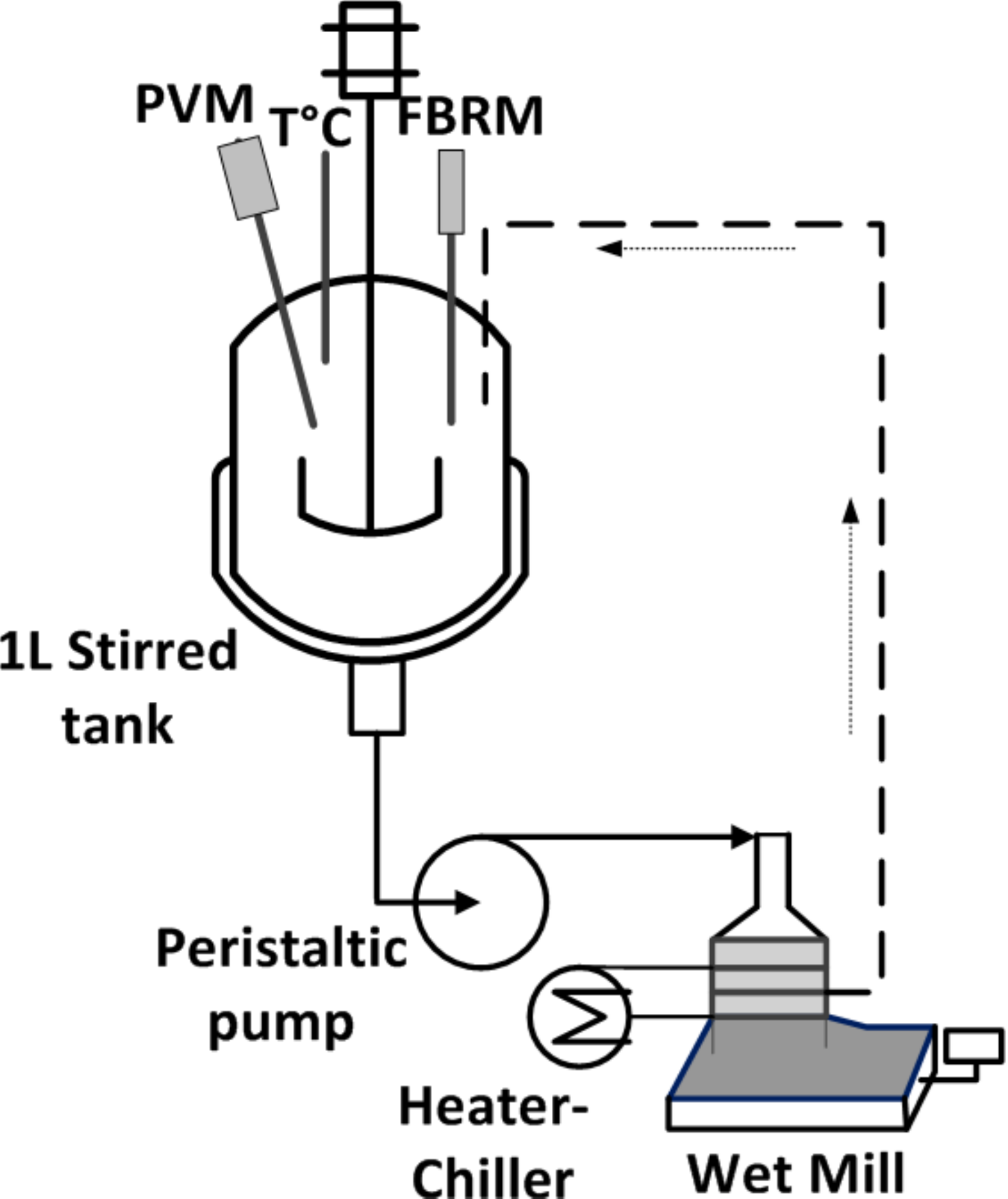}}
\caption{Sketch of the setup used for the wet milling processes in this study.}
\label{fig1}
 \end{figure}

The experiments were conducted in a closed loop consisting of a Mettler Toledo OptiMax Workstation made up of a 1L stirred tank crystallizer equipped with an inline Hastelloy Pt100 temperature sensor. The Workstation was connected to a Watson Marlow Du520 peristaltic pump, which was subsequently connected to an IKA MagicLab (module UTL) rotor stator wet mill. The wet mill was finally connected back to the Workstation to close the loop as sketched in Fig. \ref{fig1}. The temperature of the wet mill was controlled with a Lauda heater/chiller unit as shown in Fig. \ref{fig1}. The process conditions of temperature and stirring speed of the slurry in the Workstation were controlled using the iControl v5.2 software from Mettler Toledo.

Data related to the size and the shape of the particles in the wet milling processes was captured with the Mettler Toledo FBRM G400 series and PVM V819  sensors within the stirred tank as sketched in Fig. \ref{fig1}. The FBRM sensor produces a narrow laser beam which moves in a circular trajectory. The beam, when incident on a particle, traces out a chord on the particle. The lengths of chords measured over a pre-set period of time for particles in the slurry is then reported as a chord length distribution (CLD) \cite{Kail2007,Kail2009,Heinrich2012,Agimelen2015}. The PVM sensor takes images of the particles using eight (some of which can be switched off) laser beams. The images are recorded on a CCD array which are subsequently transferred to a computer. The size of each image of the PVM V819 is $1360\times 1024$ pixels with a pixel size of $0.8\mu$m \cite{Agimelen2016}. 

Offline particle size and shape analyses were carried out using the Malvern Morphologi G3 instrument. The Morphologi instrument consists of a dispersion unit which utilizes compressed air to disperse the particles over a glass plate. Images of particles on the plate are captured using a camera with a microscope lens. The images are then analysed by the instrument software for size and shape information. 

\subsection{Experimental procedure}
\label{subsec2-3}

At the start of each experiment, a saturated solution of approximately 900ml was generated inside the OptiMax vessel at 25$^{\circ}$C by adding the required quantity of solid.
 The temperature was ramped to 40 - 50$^{\circ}$C to speed up dissolution and subsequently cooled to 25$^{\circ}$C over 20min. Once the temperature had reached the setpoint value, some solid particles (whose mass varied with the different materials) was added and allowed to equilibrate for 60min. Before the addition of these solid particles, a sample of the original material (starting material) was initially analysed with the offline Morphologi instrument for PSD and aspect ratio information. 
 
 After the equilibration period (covering a period $T_1$), the peristaltic pump and wet mill were started simultaneously. The speed of the pump was maintained at 50rpm throughout the experiments while that of the wet mill was initially set to 6000rpm (for a duration $T_2$) after which it was increased in stages. At the next stage (with duration $T_3$) of the process, the speed of the wet mill was increased to 10,000rpm, and subsequently to 14,000rpm (for a duration $T_4$) and finally to 18,000rpm (for a duration $T_5$). The temperature of the mill outlet was regulated manually by adjusting the heater chiller setpoint in order to maintain it at 25$^{\circ}$C and prevent dissolution.  The time intervals $T_1$ to $T_5$ varied from 30 - 90min for each material. 
 
At the end of the time interval $T_5$, the suspension was filtered and washed in a Buchner funnel. The same solvents which were used in the experiments for benzoic acid and metformin hydrochloride were used for washing each material at the end of their respective experiments, while paracetamol was washed with chilled water. Each of the cakes obtained at the end of each wet milling process was dried overnight in a vacuum oven. Like  the starting material, samples of the dry cakes (milled product) obtained at the end of each wet milling process were analysed for PSD and aspect ratio information using the offline Morphologi instrument.

\subsubsection{Benzoic acid}
\label{subsubsec2-3-1}

Benzoic acid particles were prepared by using antisolvent crystallisation (after dissolution of the original sample from Sigma-Aldrich) in order to obtain long needle shaped crystals. The particles were filtered and dried before being suspended in water for the milling experiment. The particles were suspended in water (saturated with benzoic acid) due to the low solubility of benzoic acid in water. However, due to poor wettability of benzoic acid in water, the surfactant Tween 20 was used at a concentration of 2ml/L. The solid loading in this experiment was 1.6\% w/w.

\subsubsection{Paracetamol}
\label{subsubsec2-3-2}

The original paracetamol sample from Sigma-Aldrich was dissolved in isoamyl alcohol after which prism like particles were obtained by cooling crystallisation. The particles obtained from the cooling crystallisation were then suspended in a saturated solution of paracetamol in 2-propanol for the wet milling experiment. The solid loading in this case was 4.2\% w/w. Although the solubility of paracetamol in 2-propanol is relatively high, the solvent was chosen to avoid agglomeration.

\subsubsection{Metformin hydrochloride}
\label{subsubsec2-3-3}

The metformin sample from Molekula was used directly as the particles were already needle shaped. The particles were then suspended in a saturated solution of metformin in 2-propanol (in which metformin has a low solubility and good dispersion) for the wet milling process. The solid loading in this case was 3.5\% w/w. The wet milling process for metformin was stopped at the stage $T_4$ (with the mill speed of 14,000 rpm) as the particles were quickly broken in this case.

\subsection{Data analysis}
\label{subsec2-4}

As mentioned in section \ref{subsec2-3}, the starting material and the milled product for each material were analysed for PSD and aspect ratio information using the offline Morphologi instrument.  The CLD data acquired using the inline FBRM sensor were analysed using a previously developed algorithm \cite{Agimelen2015} for PSD and aspect ratio information. Similarly, the images captured using the inline PVM sensor were analysed using a previously developed \cite{Agimelen2016} image processing algorithm also for size and shape information.

\subsubsection{Estimating relative number of particles}
\label{subsubsec2-4-1}

The number of particles produced during each wet milling process relative to the number of initially suspended particles (for each of the materials) can be estimated from analysis of images and CLD data. The number of initially suspended particles can be estimated from the mass of initially suspended particles and the volume based PSD of each starting material estimated with the offline Morphologi instrument. Even though the sample of each starting material initially analysed with the offline Morphologi instrument is not the same as the sample that was suspended for each wet milling process, the estimated number of initially suspended particles will still be reasonable, as long as the particles in the original powder of each material were well mixed.

To estimate the number of initially suspended particles, the particle length $L$ is discretised and classified into $N$ bins with the characteristic length $\overline{L}_i = \sqrt{L_iL_{i+1}}$ of bin $i$ representing the length of particles in bin $i$, where $L_i$ and $L_{i+1}$ are the bin boundaries of bin $i$. The number of particles $N_i$ in bin $i$ is given as
\begin{equation}
N_i = \frac{\tilde{m}_iM_0}{\rho v_i}.
\label{eq1}
\end{equation}
Where, $\tilde{m}_i$ is the mass fraction of the particles in bin $i$, $M_0$ is the mass of the initially suspended particles, $\rho$ is the density of particles and $v_i$ is the volume of the particles in bin $i$. Approximating the shape of all particles in each bin with an ellipsoid of semi-major axis length $\overline{a}_i = \overline{L}_i/2$ and two equal semi-minor axis length $\overline{b}_i = \overline{r}_i\overline{a}_i$, where $\overline{r}_i$ is the mean aspect ratio of all the particles in bin $i$, gives the volume of the particles in bin $i$ as $v_i = \pi\overline{r}_i^2\overline{L}_i^3/6$. Since all particles have the same density, the mass fraction of the particles in bin $i$ can be replaced by their volume fraction $\tilde{v}_i$. Then the number of particles in bin $i$ becomes
\begin{equation}
N_i = \frac{6\tilde{v}_iM_0}{\rho\pi\overline{r}_i^2\overline{L}_i^3},
\label{eq2}
\end{equation}
and the number of initially suspended particles $\mathcal{N}$ is given as
\begin{equation}
\mathcal{N}=\sum_{i=1}^{N}{N_i}.
\label{eq3}
\end{equation}

The number of particles in the slurry can be estimated from CLD data using Eqs. \eqref{eq2} and \eqref{eq3}. However, since Eq. \eqref{eq2} requires the volume fraction of particles, the volume based PSD is first estimated from the CLD data. This is accomplished using the inversion algorithm developed in \cite{Agimelen2015}. Then the volume fraction of particles can be estimated, and hence the number of particles $\mathcal{N}_{CLD}$ in the slurry using Eq. \eqref{eq3}. Subsequently, the number of particles in the slurry relative to the number of initially suspended particles $\hat{N}_{CLD} = \mathcal{N}_{CLD}/\mathcal{N}$ can be estimated.

In the case of image analysis, the mean number of objects per frame $\mathcal{N}_{IMG}$ is estimated by first counting all objects which were detected in focus and contained wholly within the image frames. This number of objects is then divided by the number of frames containing at least one object in focus to obtain $\mathcal{N}_{IMG}$. Subsequently, the number of particles in images relative to the number of initially suspended particles $\hat{N}_{IMG} = \mathcal{N}_{IMG}/\mathcal{N}$ is estimated.

 \begin{figure}[tbh]
\centerline{\includegraphics[width=\textwidth]{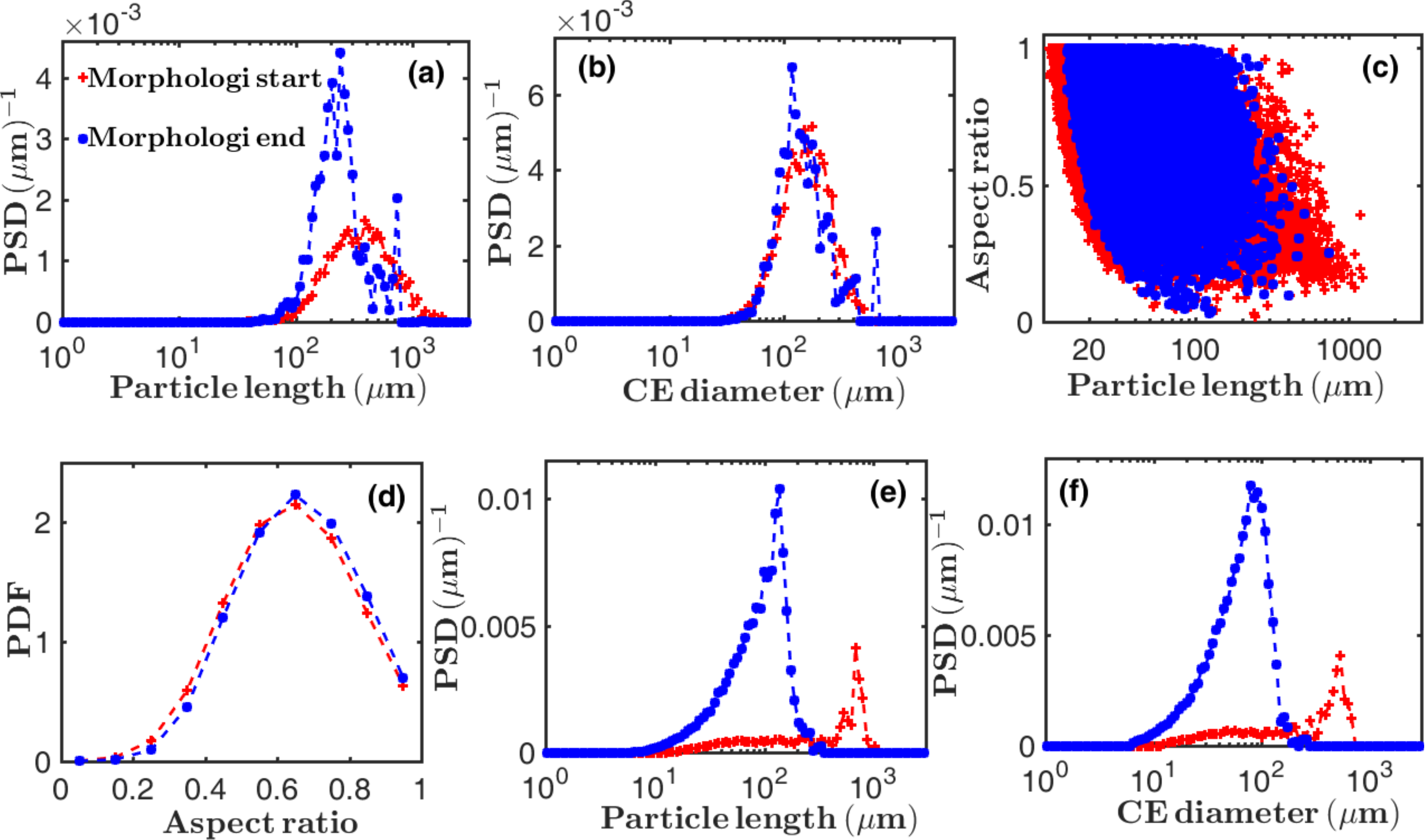}}
\caption{Volume based particle size distributions (PSDs) as functions of particle length (a) and circular equivalent (CE) diameter (b) estimated with the offline Morphologi instrument for the starting material (indicated as `Morphologi start' by the red crosses in (a)) and milled product (indicated as `Morphologi end' by the blue filled symbols in (a)). Distribution of aspect ratio with particle length (c) and the estimated probability density function (PDF) (d) of aspect ratio (both obtained using the offline Morphologi instrument) for the starting material and milled product of the benzoic acid sample. Similar to (a) and (b) are the volume based PSDs for the starting material and milled product for the paracetamol sample obtained with the offline Morphologi instrument in (e) and (f). The symbols in (b) - (f) correspond to those of the legend in (a).}
\label{fig2}
 \end{figure}
 
 \begin{figure}[tbh]
\centerline{\includegraphics[width=\textwidth]{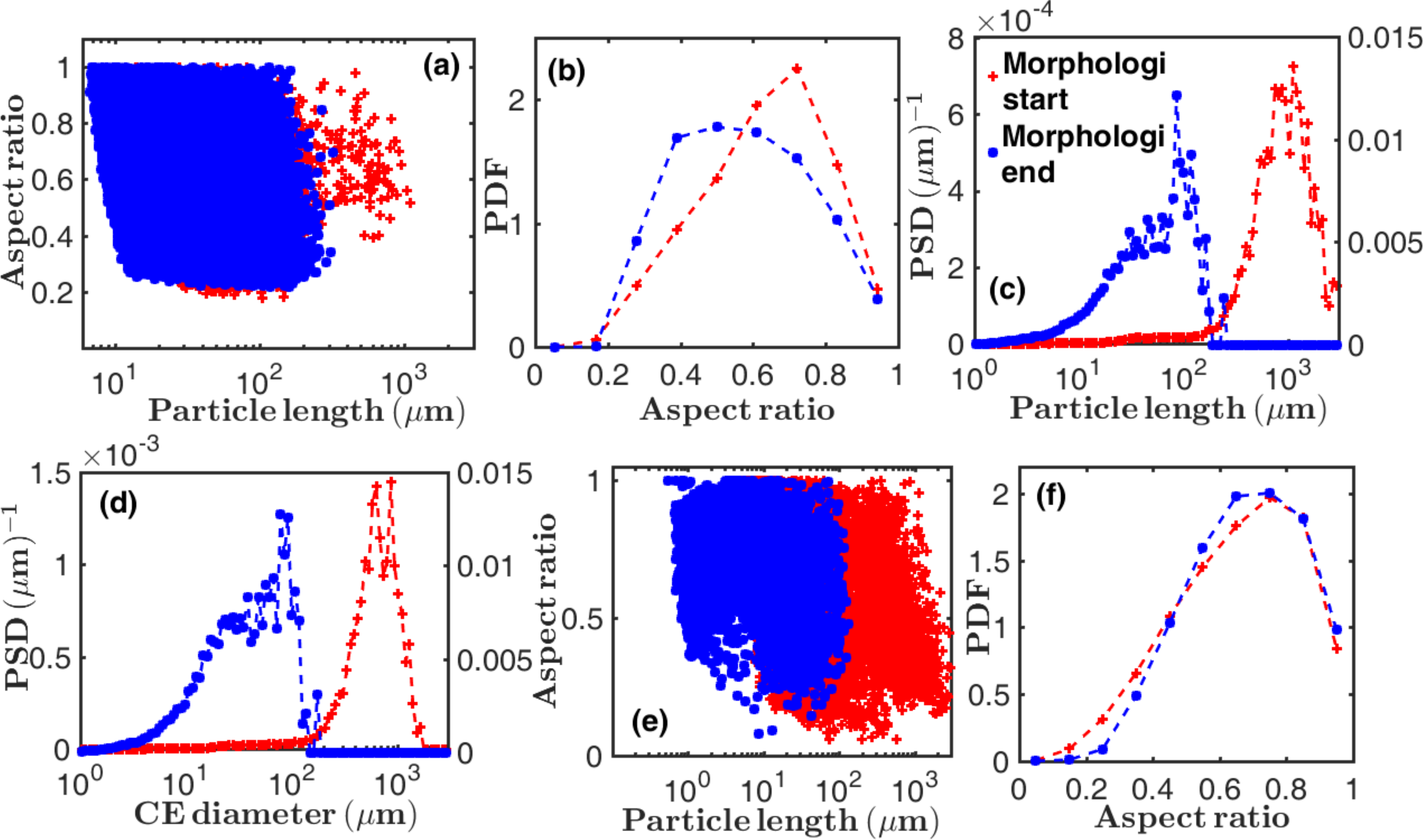}}
\caption{The distribution of aspect ratio with particle length (a) and estimated aspect ratio PDF (b) for the starting material and milled product of the paracetamol sample. Estimated volume based PSDs as functions of particle length (c) and CE diameter (d) of the starting material and milled product for the metformin sample. The data in (e) and (f) are similar to those of (a) and (b) but for the metformin sample. All data in (a) - (f) were obtained with the offline Morphologi instrument, and the symbols in (a) - (f) correspond to the legend in (c).}
\label{fig3}
 \end{figure}

 \begin{figure}[tbh]
\centerline{\includegraphics[width=0.9\textwidth]{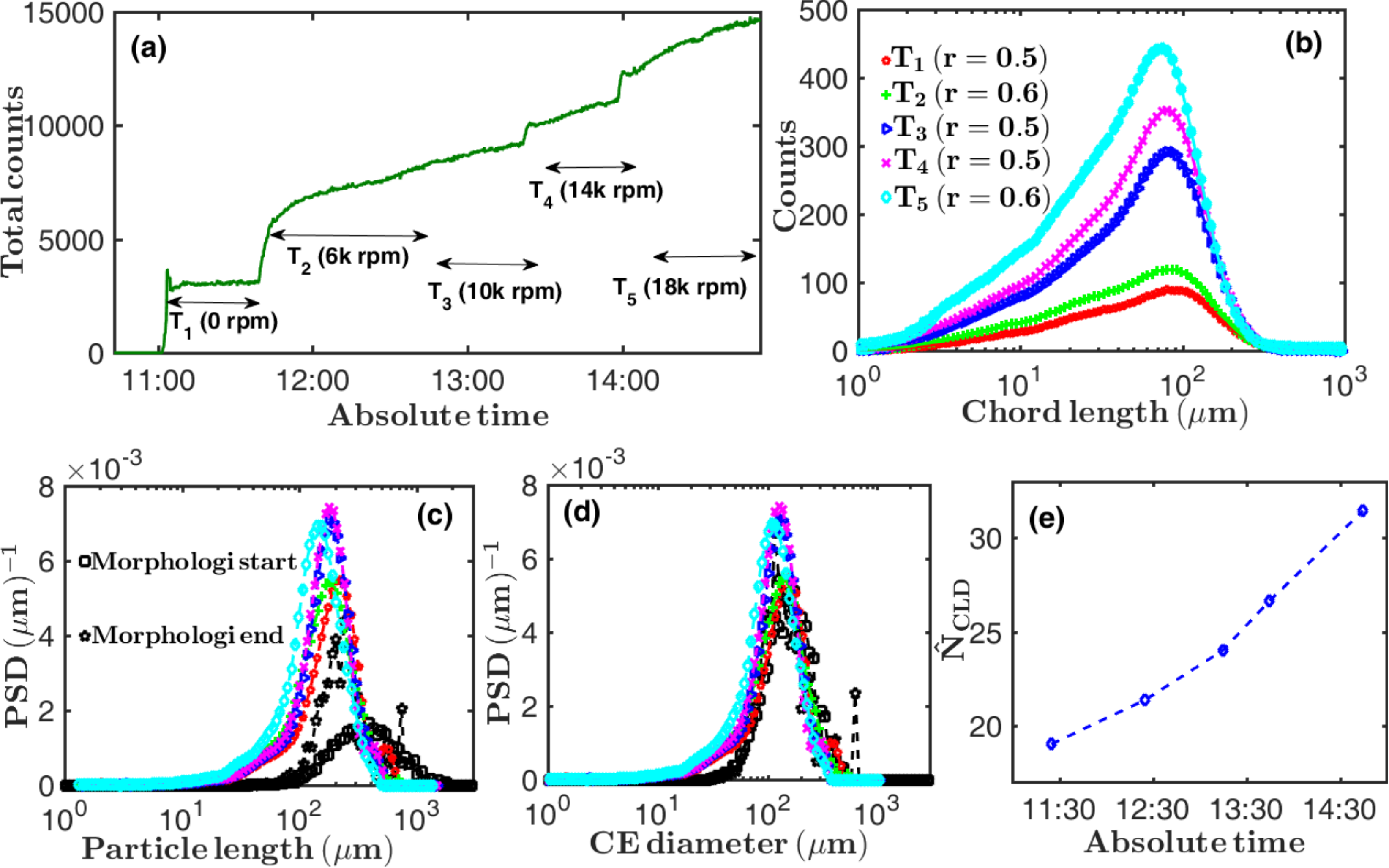}}
\caption{Total chord length counts (a) for the benzoic acid sample during the wet milling stages covering the time intervals $T_1$ to $T_5$ as indicated in the Fig. The mean chord length distributions (CLDs) acquired in the last 10mins of each stage $T_1$ to $T_5$ are shown by the symbols in (b).   The solid lines in (b) (with colours corresponding to the symbols) are the estimated CLDs at the aspect ratios $r$ indicated in the Fig. The black squares and pentagrams are the volume based PSDs as functions of particle length (c) and CE diameter (d) estimated for the starting material (indicated as Morphologi start) and milled product (indicated as Morphologi end) using the offline Morphologi instrument for the benzoic acid sample. The estimated volume based PSDs from the CLDs shown in (b) at the  aspect ratios indicated in (b) are also shown in (c) and (d). The symbols in (c) and (d) correspond to those in (b). The number of particles in the slurry relative to the number of initially suspended particles estimated from the volume based PSDs in (c) is shown in (e).}
\label{fig4}
 \end{figure}

\section{Results and discussions}
\label{sec3}

The results obtained from the analyses of data captured with the offline Morphologi instrument and as well those acquired inline (using the FBRM and PVM sensors) during the wet milling processes are discussed in subsections \ref{subsec3-1} to \ref{subsec3-3}. 

\subsection{Analysis with the offline Morphologi instrument}
\label{subsec3-1}

The volume based PSDs estimated for both the starting material and milled product for benzoic acid using the offline Morphologi instrument are shown in Figs. \ref{fig2}(a) and \ref{fig2}(b) while the distribution of aspect ratio for the same samples obtained with the same instrument are shown in Figs. \ref{fig2}(c) and \ref{fig2}(d). The data clearly show particle breakage as the peak and right tail of the volume based PSD moves to the left from the starting material to the milled product in Figs. \ref{fig2}(a) and \ref{fig2}(b). The reduction of particle length of the milled product when compared to that of the starting material in Fig. \ref{fig2}(c) also suggests particle breakage. The estimated probability density function (PDF) of aspect ratio (obtained with the offline Morphologi instrument) of the milled product of benzoic acid is very similar to that of the starting material as seen in Fig. \ref{fig2}(d). This could be because this aspect ratio PDF is number based, and hence more sensitive to fines which may be more rounded.

Similar results are obtained for paracetamol. The reduction of particle sizes as a result of the wet milling process can be seen in the left shift of the volume based PSD in Figs. \ref{fig2}(e) and \ref{fig2}(f), and the reduction in particle lengths seen in Fig. \ref{fig3}(a). This is similar to the case of benzoic acid in Figs. \ref{fig2}(a) - \ref{fig2}(c). However, the left shift of the volume based PSD for paracetamol is more pronounced than in the case of benzoic acid. Similarly, there is a more pronounced shift of the aspect ratio PDF in the case of paracetamol (in Fig. \ref{fig3}(b)) when compared with that of benzoic acid in Fig. \ref{fig2}(d). This could be due to differences in material properties between benzoic acid and paracetamol making them respond differently to the wet mill. Differences in agglomeration of the starting materials, de-agglomeration during agitation and milling and re-agglomeration during filtration and drying of the milled product could have also contributed to the differences. See section .... of the supplementary information for sample images of the starting materials and milled products for benzoic acid, paracetamol and metformin.

The situation with metformin is similar to those of benzoic acid and paracetamol. However, the left shift (due to breakage) in the volume based PSD is more significant (as seen in Figs. \ref{fig3}(c) and \ref{fig3}(d)) than in the previous two cases. This is also reflected in the shift in particle length on moving from the starting material to the milled product seen in Fig. \ref{fig3}(e). This is because of the brittle nature of metformin which resulted in most of the large particles being broken down to fines at stage $T_1$ of the wet milling process. However, as the aspect ratio PDF is number based, it does not show much of a shift on moving from the starting material to the milled product in Fig. \ref{fig3}(f).

\subsection{Analysis of CLD data}
\label{subsec3-2}

The total CLD counts at the different time intervals $T_1$ to $T_5$ during the wet milling of benzoic acid are shown in Fig. \ref{fig4}(a), while the mean CLD captured in the last 10 mins of each time interval $T_1$ to $T_5$ are shown by the symbols in Fig. \ref{fig4}(b).  The increase in total chord counts over the wet milling stages $T_1$ to $T_5$ seen in Fig. \ref{fig4}(a) clearly suggests breakage of particles during the process as the process conditions were such that there was no nucleation or growth of particles. This increase in chord counts is also seen in the increase in the peaks of the mean CLDs shown in Fig. \ref{fig4}(b).

The solid lines (with the colours corresponding to the symbols) in Fig. \ref{fig4}(b) are the estimated CLDs obtained by solving the associated inverse problem. They show near perfect agreement with the corresponding experimental data. This involves searching for a PSD at different aspect ratios\footnote{The ratio of width to length of particles} $r$ (all particles are assumed to have the same mean aspect ratio) whose corresponding CLD gives the best fit to the measured CLD \cite{Agimelen2015}. In the case of Fig. \ref{fig4}(b), these best fits were obtained at $r=0.5$ ($T_1$), $r=0.6$ ($T_2$), $r=0.5$ ($T_3$), $r=0.5$ ($T_4$) and $r=0.6$ ($T_5$) as indicated in the Fig. These values of aspect ratios for obtaining the best fit to the experimentally measured CLD are close to the peak of the estimated aspect ratio PDF obtained with the offline Morphologi instrument shown in Fig. \ref{fig2}(d). 

The PSDs estimated from the CLDs in Fig. \ref{fig4}(b) (at the best fit values of $r$) are shown in Fig. \ref{fig4}(c) (as functions of particle length) and \ref{fig4}(d) (as functions of circular equivalent (CE) diameter) for benzoic acid. The estimated PSDs (Figs. \ref{fig4}(c) and \ref{fig4}(d)) both show breakage of particles moving from $T_1$ to $T_5$. That is, the peaks of the distributions shift to the left and both the right and left tails of the distributions shift to the left on moving from $T_1$ to $T_5$. 

The main peak of the volume based PSD estimated from the CLD at $T_1$ is shifted to the left of the peak of the volume based PSD of the starting material estimated with the offline Morphologi instrument in terms of particle length in Fig. \ref{fig4}(c), but it agrees very well with the corresponding estimate for the starting material in terms of CE diameter in Fig. \ref{fig4}(d). Also, the peak of the volume based PSD estimated from the CLD at $T_5$ is shifted to the left of the estimated volume based PSD of the milled product in terms of particle length in Fig. \ref{fig4}(c), but shows a better agreement with the estimated PSD of the milled product in terms of CE diameter in Fig. \ref{fig4}(d). However, the volume based PSD estimated at $T_1$ has a minor peak at a particle  length of about $500\mu$m (Fig. \ref{fig4}(c)) which is to the right of the peak (at a particle length of about $300\mu$m in Fig. \ref{fig4}(c)) of the volume based PSD estimated for the starting material using the offline Morphologi instrument.

The occurrence of the minor peak at a particle length of about $500\mu$m in the estimated volume based PSD from the CLD at $T_1$, and fat right tail (extending to particle length of 2000$\mu$m) of the estimated volume based PSD of the starting material obtained with the offline Morphologi instrument suggest the presence of particles of length of up to about $2000\mu$m in the starting material. These length dimensions appear not to have been captured very well in the CLD data. Part of the reason could be that the circular laser beam (with a diameter of $5.3\,$mm for the FBRM G400 used in this work) of the FBRM sensor has a much lower probability of capturing this length dimension than that predicted by the ideal CLD model used in this work. This ideal CLD model \cite{Li2005n1} assumes among other things that all particles lie on the focal plane of the laser spot of the FBRM instrument, and that the laser spot makes a straight chord on the particles. However, as the length dimensions of the particles become comparable to the diameter of the laser beam, the curvature of the chord becomes more pronounced. Then the estimated probability of obtaining a chord of a given length becomes less accurate. Particles detected out of focus also contribute to this inaccuracy. Another possible reason why the length dimensions of around $200\mu$m were not captured well in the CLD data could be that some of the particles in the starting material had been broken down during the time interval $T_1$ reducing their contribution to the CLD count. 

The estimated volume based PSD from the CLD at $T_5$ shows a higher proportion of particles of length between about $20\mu$m to about $70\mu$m in Fig. \ref{fig4}(c) (or CE diameter of about $20\mu$m to about $50\mu$m in Fig. \ref{fig4}(d)) when compared with the estimated volume based PSD of the milled product using the offline Morphologi instrument. A possible reason for this discrepancy could be the large amount of bubbles produced during the wet milling of the benzoic acid sample. These bubbles lead to chord splitting \cite{Kail2007,Heinrich2012} at their boundaries as they are transparent to the FBRM laser. This effect leads to an unphysical high count of short chords, and thereby leads to an over estimation of fines in the estimated volume based PSD. It could also be that there was significant agglomeration of the milled product during filtration and drying, thereby leading to a higher count of large particles which dominate the estimated volume based PSD obtained with the offline Morphologi instrument.

The number of particles in the slurry relative to the number of initially suspended particles $\hat{N}_{CLD}$ estimated from the CLD data is shown in Fig. \ref{fig4}(e). The estimate is made using the volume based PSD shown in Fig. \ref{fig4}(c) (at $T_1$ to $T_5$) in Eq. \eqref{eq2}. The data shows an increase in the number of particles in the slurry which is to be expected in a wet milling process.

 \begin{figure}[tbh]
\centerline{\includegraphics[width=\textwidth]{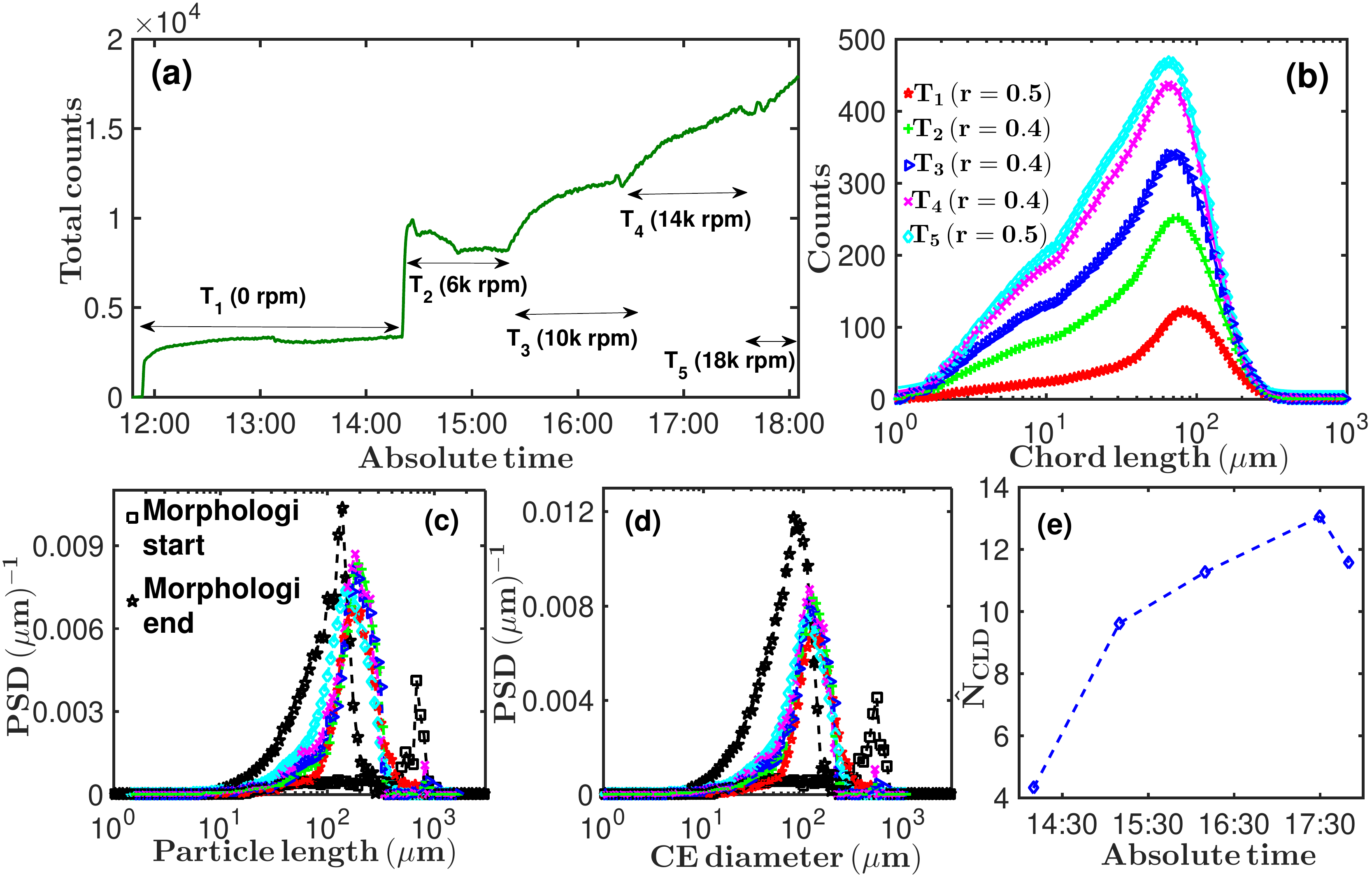}}
\caption{Total CLD counts (a), mean and estimated CLDs (b), estimated volume based PSDs (c) and (d), and relative number of particles (e) for the paracetamol sample similar to Fig. \ref{fig4} for benzoic acid. }
\label{fig5}
 \end{figure}

The increase in total CLD counts over the stages $T_1$ to $T_5$ of the wet milling process for paracetamol in Fig. \ref{fig5}(a) suggests particle breakage. This is similar to the case of benzoic acid in Fig. \ref{fig4}(a). The mean of the measured CLD in the last 10 mins of each of the stages $T_1$ to $T_5$ of the wet milling process for the paracetamol sample is shown by the symbols in Fig. \ref{fig5}(b). The solid lines (with colours corresponding to the symbols) in the same Fig. are the estimated CLDs at aspect ratios  $r=0.5$ ($T_1$), $r=0.4$ ($T_2$), $r=0.4$ ($T_3$), $r=0.4$ ($T_4$) and $r=0.5$ ($T_5$). 
 The aspect ratio of $r=0.5$ estimated at $T_1$ is slightly less than the position of the peak (at about $r=0.7$ in Fig. \ref{fig3}(b)) of the PDF of the  aspect ratio obtained using the offline Morphologi instrument for the starting material of paracetamol. However, this PDF covers a broad range from about $r=0.2$ to $r=1$. This suggests that the mean aspect ratio of $r=0.5$ at $T_1$ obtained from the CLD is a reasonable value. A similar situation holds for the milled product of paracetamol seen in Fig. \ref{fig3}(b) obtained with the offline Morphologi instrument. The estimated aspect ratios cover a broad range from about $r=0.2$ to $r=1$ (with a peak close to $r=0.5$). This suggests that the mean aspect ratio of $r=0.5$ estimated from the CLD data at $T_5$ is reasonable. Similar to the case of benzoic acid, breakage of particles is reflected in an increase in total CLD counts for the paracetamol sample as seen in Fig. \ref{fig5}(a). This breakage also leads to an increase in the peaks of the mean CLDs collected in the last 10 mins of each time intervals $T_1$ to $T_5$ as shown by the symbols in Fig. \ref{fig5}(b), where the solid lines with corresponding colours to the symbols are the estimated CLDs at the aspect ratios indicated in the Fig.

Furthermore, the estimated PSDs (from the CLDs in Fig. \ref{fig5}(b)) in Fig. \ref{fig5}(c) (as functions of particle length) and Fig. \ref{fig5}(d) (as functions of CE diameter) both show particle breakage moving from $T_1$ to $T_5$. The peaks of the estimated PSDs from the CLDs from $T_1$ to $T_5$ show a slight drift to the left, however, a stronger signature of particle breakage is seen in the tails of the estimated PSDs from $T_1$ to $T_5$ as seen in Figs. \ref{fig5}(c) and \ref{fig5}(d).

The volume based PSD estimated from the starting material using the offline Morphologi instrument shows peaks at a particle length of $800\mu$m (Fig. \ref{fig5}(c)) and at a CE diameter of $500\mu$m (Fig. \ref{fig5}(d)). These peaks are far to the right of the peaks of the estimated volume based PSD from the CLD at $T_1$ which occur at a particle length of about $200\mu$m (Fig. \ref{fig5}(c)) and a CE diameter of about $100\mu$m (Fig. \ref{fig5}(d)). The reason for this huge discrepancy in the estimated PSDs from CLD and the offline Morphologi instrument could be because a significant number of particles may have been agglomerated during the offline measurement of the starting material. These agglomerates may have de-agglomerated upon agitation in the stirred tank during the time interval $T_1$, hence these large dimensions were not captured in the CLD data.

However, the peaks of the volume based PSD estimated from the CLD at $T_5$ are closer (although shifted to the right) to those estimated for the milled product using the offline Morphologi instrument as seen in Figs. \ref{fig5}(c) and \ref{fig5}(d).  The estimated volume based PSD for the milled product obtained with the offline Morphologi instrument suggests a higher proportion of particles of length less than about $100\mu$m in Fig. \ref{fig5}(c) and CE diameter less than about $70\mu$m (Fig. \ref{fig5}(d)) when compared with similar estimates from the CLD at $T_5$. Part of the reason for the discrepancy could be the production of small particles during filtration and drying due to breakage.

Similar to the case of benzoic acid in Fig. \ref{fig4}(e), the number of particle relative to the number of initially suspended particles $\hat{N}_{CLD}$ for the paracetamol sample shows an increase in particle number over the course of the wet milling process as seen in Fig. \ref{fig5}(e). However, there is a slight decrease in $\hat{N}_{CLD}$ after 17:30 in Fig. \ref{fig5}(e). This is because the left shift in the estimated volume based PSD (in Fig. \ref{fig5}(c)) is not large enough to counter the effect of the larger aspect ratio ($r=0.5$) estimated at $T_5$ for paracetamol. This causes a decrease in the number of particles estimated with Eq. \eqref{eq2}.

 \begin{figure}[tbh]
\centerline{\includegraphics[width=\textwidth]{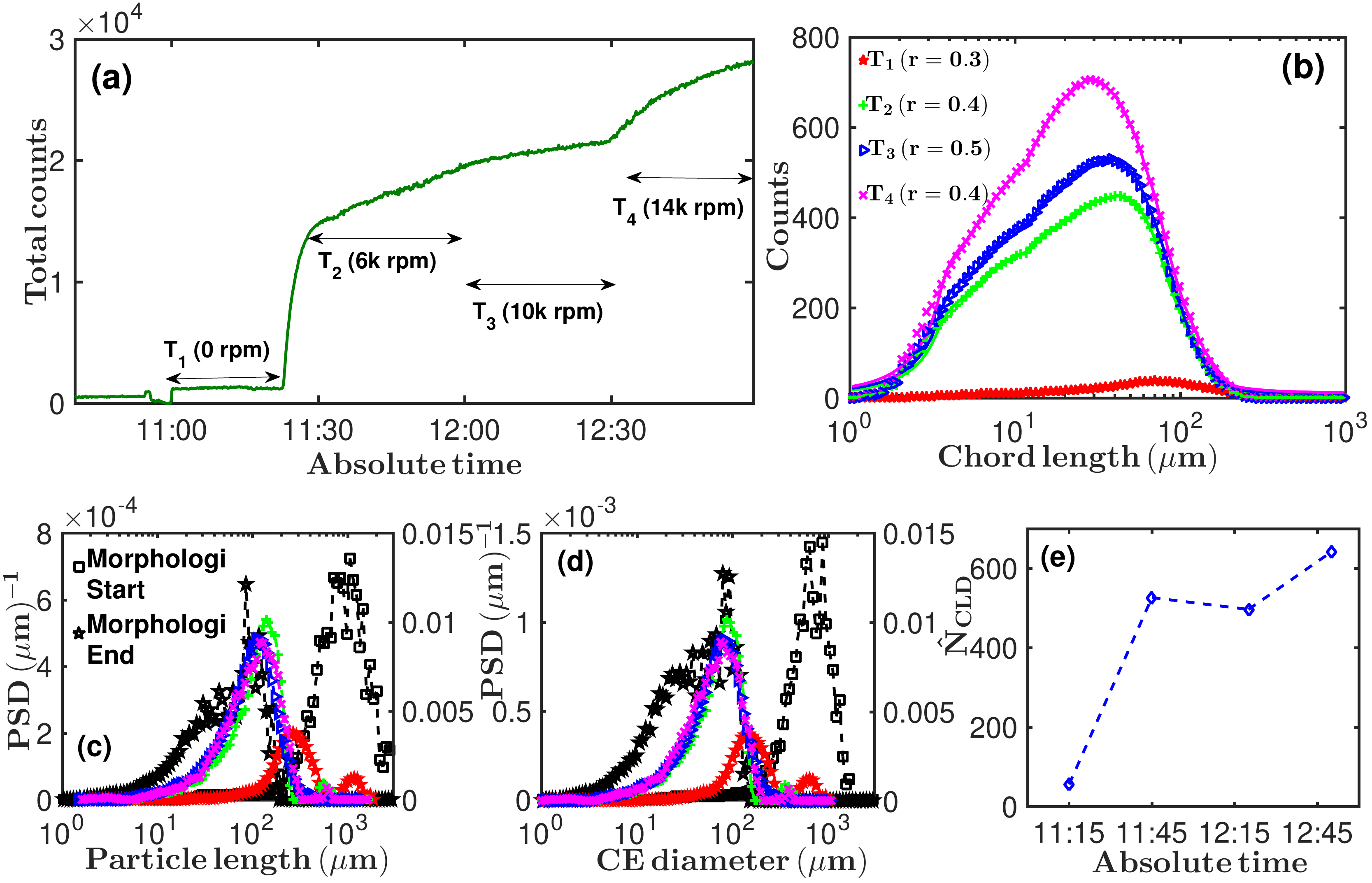}}
\caption{Measured and estimated quantities for the wet milling of metformin similar to Fig. \ref{fig4}.}
\label{fig6}
 \end{figure}

Similar to the cases of benzoic acid and paracetamol, breakage of particles during the wet milling of the metformin sample is reflected in the increase in total CLD counts in Fig. \ref{fig6}(a) and the mean CLDs measured in the last 10 mins of each stage $T_1$ to $T_4$ shown by the symbols in Fig. \ref{fig6}(b). This is also seen in the increase in the number of particles relative to that of initially suspended particles $\hat{N}_{CLD}$ in Fig. \ref{fig6}(e). As mentioned in subsection \ref{subsubsec2-3-3}, the stage $T_5$ of the wet milling process was not conducted for the metformin sample, as the particles had broken down so much at the later part of $T_4$ that there was very little contrast between the background and the objects in the PVM images collected. See the supplementary information for sample images of the materials used in this work.

The estimated CLDs corresponding to the measured CLDs (symbols in Fig. \ref{fig6}(b)) at the time intervals $T_1$ to $T_4$ are shown by the solid lines in Fig. \ref{fig6}(b) at the aspect ratios indicated in the Fig.
The PVM images (section 2 of the supplementary information) collected at $T_1$ for the metformin sample show that some of the metformin particles were long rod-like objects at the start of the process. This is consistent with the mean aspect ratio of $r=0.3$ estimated from the CLD data (in Fig. \ref{fig6}(b)) at  $T_1$ for metformin. But at variance with the estimated aspect ratio PDF (shown in Fig. \ref{fig3}(f)) for metformin obtained with the offline Morphologi instrument; the starting material and milled product have peaks close to $r=0.8$. This suggests that the aspect ratio PDF estimated using the offline Morphologi instrument (which is number based) is dominated by the smaller particles (which are more likely to be rounded) in the metformin sample. This is in contrast to the CLD method which is biased towards larger particles \cite{Agimelen2015}, and in this case rod-like particles. Some sample images for both the starting material and milled product for metformin obtained with the offline Morphologi instrument in section 1 of the supplementary information show some of these small particles.

The estimated volume based PSD (from the CLD data in Fig. \ref{fig6}(b)) at the time interval $T_1$ shows a minor peak at a particle length close to 1000$\mu$m in Fig. \ref{fig6}(c) and close to a CE diameter of 700$\mu$m in Fig. \ref{fig6}(d). 
These values are close to the peaks of the volume based PSD of the starting material estimated with the offline Morphologi instrument as seen in Figs. \ref{fig6}(c) and \ref{fig6}(d). This clearly reflects the presence of the long rod-like particles (considering the aspect ratio of $r=0.3$ estimated at $T_1$) in the slurry as seen in Fig. 9 in section 2 of the supplementary information. However, the main peaks of the volume based PSD estimated from the CLD at $T_1$ occur at a particle length of about $200\mu$m (Fig. \ref{fig6}(c)) and CE diameter of about $100\mu$m (Fig. \ref{fig6}(d)). The reason why the main peak is markedly shifted to the left could be because the length dimension of up to about $3000\mu$m is not well captured by the ideal CLD model used in this work. This is similar to the case of benzoic acid discussed previously.

However, the estimated PSDs from the CLDs from $T_1$ to $T_4$ show (Figs. \ref{fig6}(c) and \ref{fig6}(d)) a significant breakage of particles in agreement with the estimated PSDs (Figs. \ref{fig6}(c) and \ref{fig6}(d)) of the milled product obtained using the offline Morphologi instrument.

 \begin{figure}[tbh]
\centerline{\includegraphics[width=0.9\textwidth]{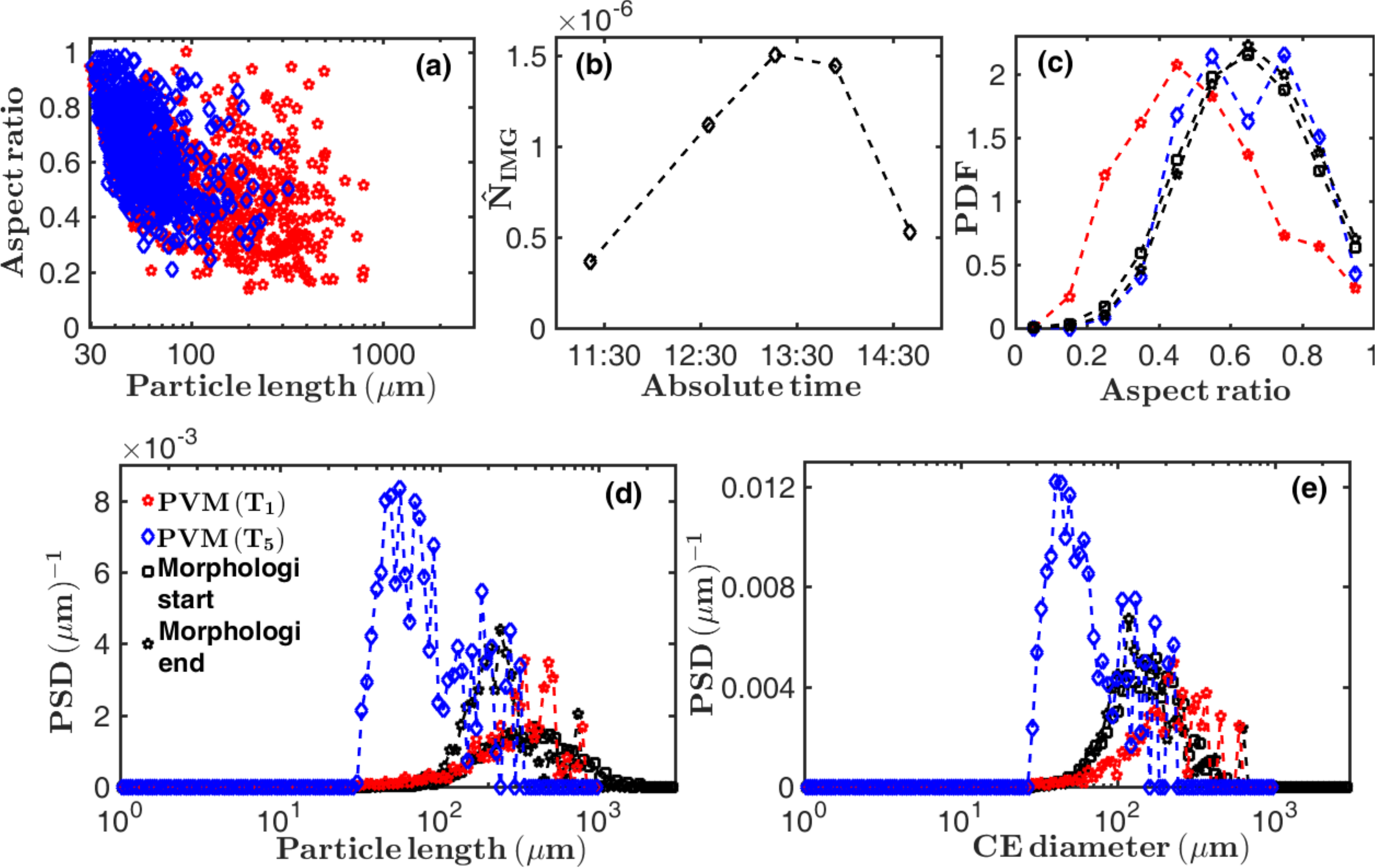}}
\caption{The scatter plot of aspect ratio as a function of particle length from the detected objects in the inline PVM images at the stages covering the time intervals $T_1$ and $T_5$ of the wet milling process of benzoic acid are shown in (a). The symbols in (a), (c), (d) and (e) are as indicated in (d). The mean number of objects per frame in PVM images relative to the number of initially suspended particles is shown in (b).
The PDF of aspect ratio estimated for the starting material and milled product using the offline Morphologi instrument are shown in (c). The corresponding estimated aspect ratio PDF from the detected objects in the PVM images at $T_1$ and $T_5$ are also shown in (c). The estimated volume based PSDs as functions of particle length (d) and CE diameter (e) for the starting material and milled product (obtained with the offline Morphologi instrument) as well as the corresponding estimates from PVM images at $T_1$ and $T_5$ are shown in (d) and (e).}
\label{fig7}
 \end{figure}

\subsection{Analysis of inline PVM images}
\label{subsec3-3}

The scatter plot of aspect ratio versus particle length (Fig. \ref{fig7}(a)) obtained from the analysis of images captured inline with the PVM sensor for the benzoic acid particles suggests breakage of the particles moving from $T_1$ to $T_5$. This is also seen in the volume based PSDs at the time intervals $T_1$ to $T_5$ estimated from the detected objects in the PVM images in the specified time interval as shown in Figs. \ref{fig7}(d) (as functions of particle length) and \ref{fig7}(e) (as functions of CE diameter). This breakage of particles is also reflected in the mean number of objects per frame $\hat{N}_{IMG}$ relative to the number of initially suspended particles seen in Fig. \ref{fig7}(b). The average number of objects increase up to about time 13:30 before decreasing. As the wet milling progresses and the particle sizes reduce, the number of images containing at least one particle wholly within the frame increases. However, most of these frames only contain a few particles in focus. Hence, even though the number of objects detected in focus increases, this increase is not enough to match the increase in the number of frames. Therefore, the value of $\hat{N}_{IMG}$ decreases after some time. In the case of benzoic acid, this decrease occurs after around the time 13:30.

However, the estimated volume based PSD from the PVM images are quite noisy as a significant number of objects are either out of focus or have sizes below the resolution limit of the PVM or touch the boundaries of the image frame. The objects which fall into the aforementioned categories are rejected by the image processing algorithm \cite{Agimelen2016}. However, the trend of particle breakage from $T_1$ to $T_5$ agrees with that of the data for the starting material and milled product obtained with the offline Morphologi instrument as seen in Figs. \ref{fig7}(d) and \ref{fig7}(e).

The particle lengths estimated from the PVM images at $T_1$ in Fig. \ref{fig7}(d) show a truncation at a length of about $800\mu$m, whereas the corresponding estimate (shown in the same Fig.) of the starting material with the offline Morphologi instrument extends to lengths larger than $1000\mu$m. Part of the reason for this discrepancy is that particles longer than about $800\mu$m (the PVM image frame has dimensions of $1088 \times 819\mu$m) have a very low chance of fitting within the image frame and hence get rejected by the image processing algorithm. Furthermore, some of the particles in the inline images may be detected partially in focus resulting in an underestimation of their lengths. There may also have been breakage of some of the long rod-like particles upon agitation in the stirred tank.

The (noisy) peak of the estimated PSD for benzoic acid from the inline PVM images at $T_1$ in Figs. \ref{fig7}(d) (as a function of particle length) and \ref{fig7}(e) (as a function of CE diameter) are close to the corresponding estimated PSDs of the starting material obtained with the offline Morphologi instrument as seen in the same Figs. In contrast, the estimated volume based PSD from the PVM images at $T_5$ has peaks that are shifted far to the left of the corresponding estimate from the milled product using the offline Morphologi instrument as seen in Figs. \ref{fig7}(d) and \ref{fig7}(e). 

Furthermore, the left tail of the estimated volume based PSD from the PVM images at $T_5$ suggest a significantly higher proportion of small particles when compared to the corresponding estimate from the milled product using the offline Morphologi instrument. This is despite the fact that the image processing algorithm rejects particles of sizes below about $30\mu$m due to the resolution limit of the PVM sensor images \cite{Agimelen2016}. This is similar to the case of the estimated volume based PSD from the CLD data for benzoic acid at $T_5$ (Figs. \ref{fig4}(c) and \ref{fig4}(d)) when compared with the volume based PSD of the milled product (Figs. \ref{fig4}(c) and \ref{fig4}(d)) estimated with the offline Morphologi instrument. Hence, it is very likely that significant agglomeration of the milled product of benzoic acid occurred during filtration and drying.

 The estimated aspect ratio PDF from the PVM images for benzoic acid at $T_1$ has a peak at $r=0.45$ (Fig. \ref{fig7}(c)) which is close to the estimated mean aspect ratio of $r=0.5$ from the CLD data for benzoic acid at $T_1$ but less than the peak value of $r=0.65$ estimated for the starting material with the offline Morphologi instrument as seen in Fig. \ref{fig7}(c). However, the estimated aspect ratio PDF from the inline PVM images at $T_5$ in Fig. \ref{fig7}(c) has two peaks at $r=0.55$ and $r=0.75$. The mean value of these aspect ratios are close to the estimated mean aspect ratio of $r=0.6$ from the CLD data for benzoic acid at $T_5$, as well as the peak position of the estimated aspect ratio PDF from the offline Morphologi instrument in Fig. \ref{fig7}(c). In this case of benzoic acid, there is more consistency in the shape information obtained by the three methods despite the inconsistencies in the PSD estimation.
 \begin{figure}[tbh]
\centerline{\includegraphics[width=\textwidth]{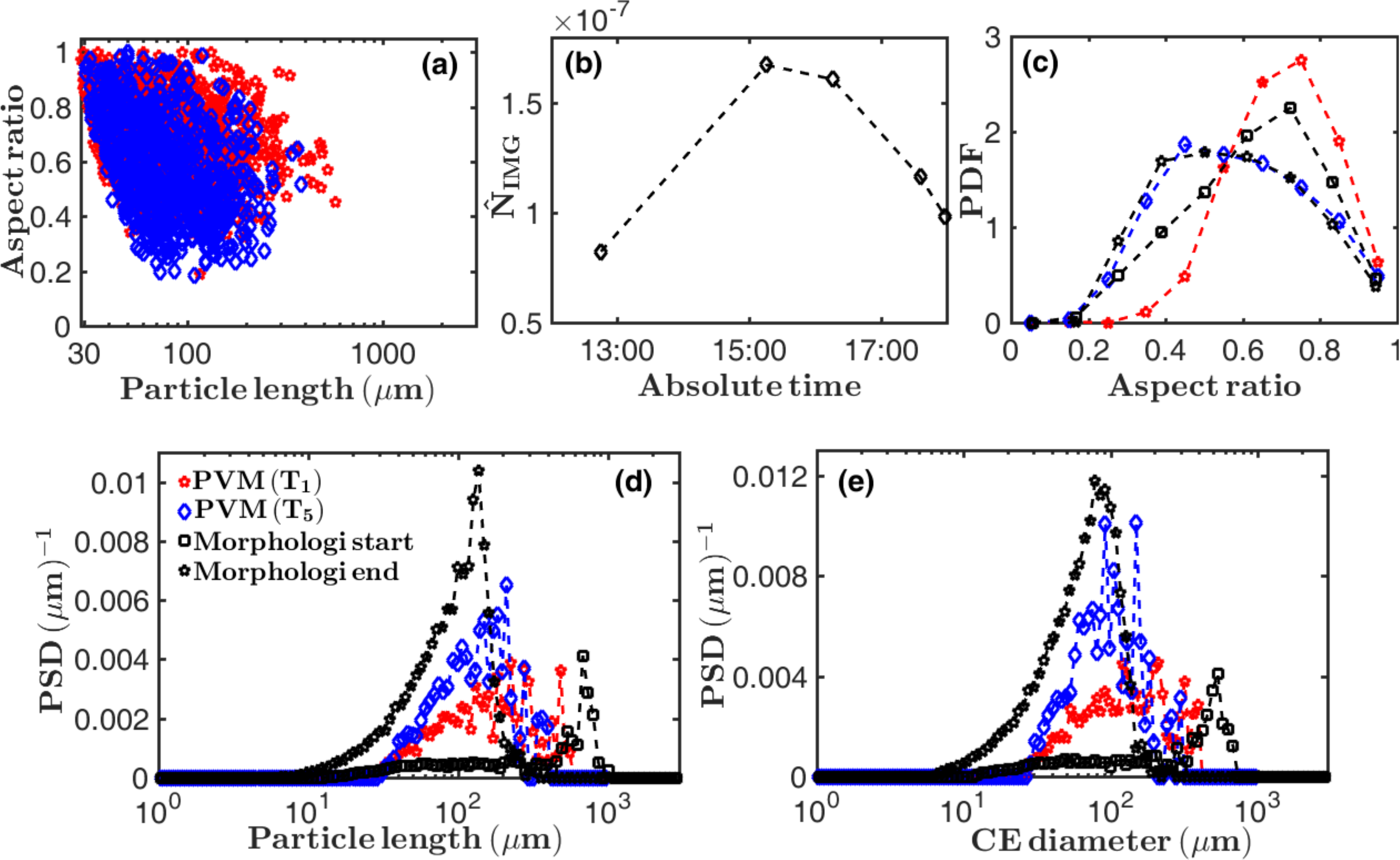}}
\caption{Estimates of aspect ratios (a) and (c), mean number of objects per frame (b) and volume based PSDs (d) and (e) for the paracetamol sample similar to Fig. \ref{fig7} for benzoic acid.}
\label{fig8}
 \end{figure}

 \begin{figure}[tbh]
\centerline{\includegraphics[width=\textwidth]{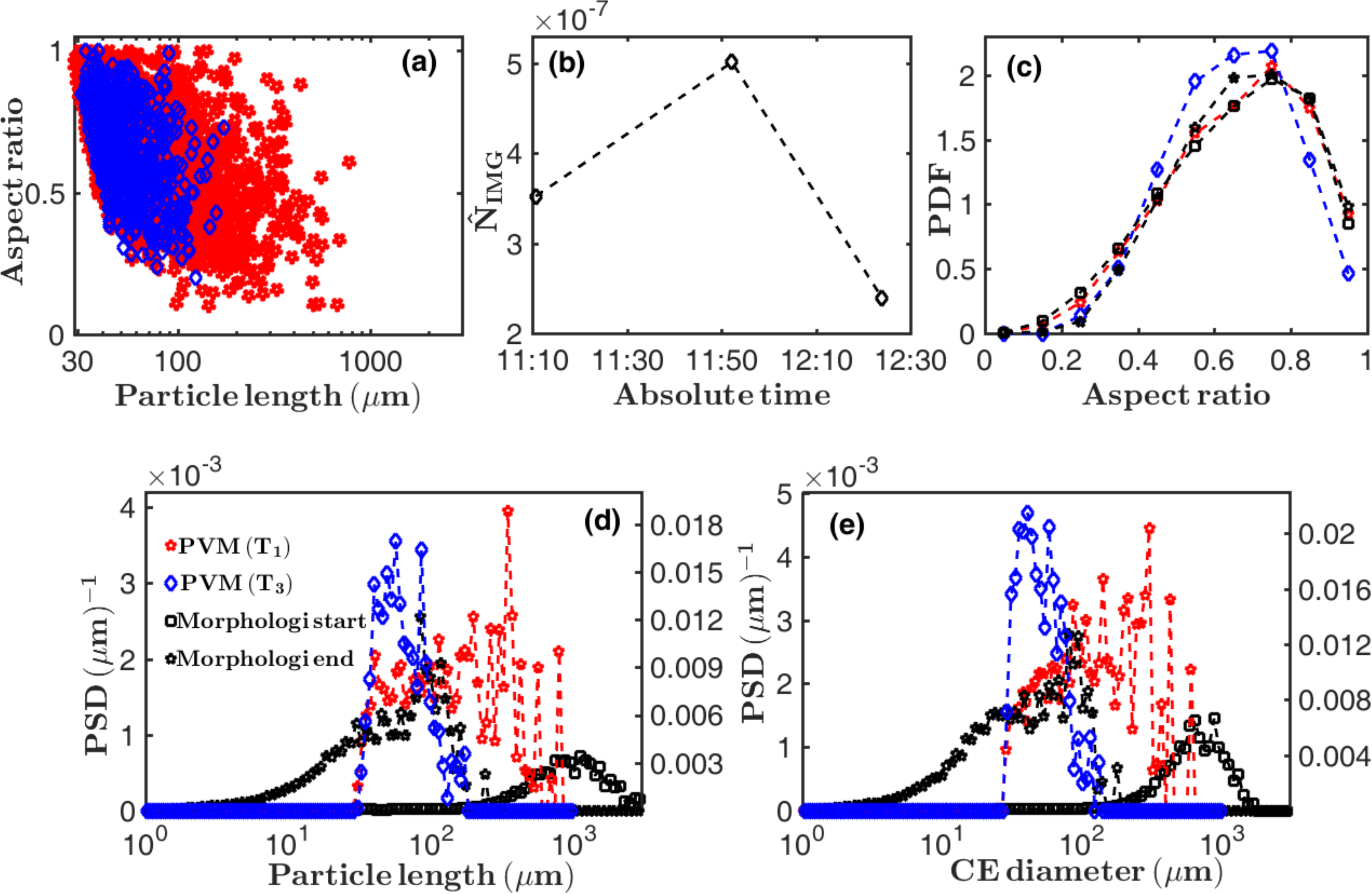}}
\caption{Estimates of quantities for the metformin sample similar to those of benzoic acid and paracetamol in Figs. \ref{fig7} and \ref{fig8} respectively.}
\label{fig9}
 \end{figure}
The data in Fig. \ref{fig8} are estimates from inline images for paracetamol similar to those of benzoic acid shown in Fig. \ref{fig7}. Breakage of particles can be seen in the scatter plot of the aspect ratio versus particle length in Fig. \ref{fig8}(a) and the estimated PSDs in Figs. \ref{fig8}(d) (as functions of particle length) and \ref{fig8}(e) (as functions of CE diameter). This breakage can also be seen in the estimated mean number of images per frame $\hat{N}_{CLD}$ relative to the number of initially suspended particles in Fig. \ref{fig8}(b). The value of $\hat{N}_{CLD}$ decreases just after time 15:00 (Fig. \ref{fig8}(b)) for similar reasons discussed earlier for the case of benzoic acid in Fig. \ref{fig7}(b).

The volume based PSD of the starting material estimated with the offline Morphologi instrument has a peak at a particle length of $800\mu$m (Fig. \ref{fig8}(d)) and CE diameter of $500\mu$m (Fig. \ref{fig8}(e)). This is far to the right of the (noisy) peaks of the volume based PSD (at $T_1$) estimated from the inline PVM images which occur at a particle length of $400\mu$m (Fig. \ref{fig8}(d)) and CE diameter of $200\mu$m (Fig. \ref{fig8}(e)). Even though larger particles have a higher chance of being rejected by the image processing algorithm due to their higher chance of making contact with the image frame, the PVM images do not show the presence of particles of lengths up to $800\mu$m at $T_1$ as seen in Fig. \ref{fig8}(a). This situation is similar to that faced when the estimated volume based PSD of the starting material (using the offline Morphologi instrument) of paracetamol was compared with that from the CLD  data for paracetamol at $T_1$. This further strengthens the suggestion that a significant number of the paracetamol particles (of the starting material) may have been agglomerated which then de-agglomerated upon loading and agitation in the stirred tank. 

However, the position of the peaks of the estimated volume based PSD from the PVM images at $T_5$ is closer to those of the milled product as seen in Figs. \ref{fig8}(d) and \ref{fig8}(e), although the estimates (using the offline Morphologi instrument) suggest a higher proportion of fines as seen in the left tails of the volume based PSD in Figs. \ref{fig8}(d) and \ref{fig8}(e). This is partly due to the resolution limit of the PVM images discussed earlier and the possible production of more fines during filtration and drying due to breakage.

Similarly, the peak of the aspect ratio PDF estimate from the PVM images at $T_1$ is very close to the corresponding estimate of the starting material using the offline Morphologi instrument as seen in Fig. \ref{fig8}(c), but the estimated estimated aspect ratio PDF (from the inline PVM images) at $T_1$ suggests a lower proportion of particles with aspect ratio $r\lesssim 0.5$. Furthermore, the estimated aspect ratio PDF from the PVM images at $T_5$ shows a very good agreement with the corresponding estimate of the milled product obtained with the offline Morphologi instrument. The level of agreement between the estimated aspect ratio PDF from the PVM images (at $T_1$ and $T_5$) and the corresponding estimates from the starting material and milled product using the offline Morphologi instrument demonstrates that more robust estimates of PSD can be made possibly from a combination of inline PVM images and CLD in a multi-objective approach.

Similar to the cases of benzoic acid and paracetamol, breakage of particles during the wet milling process for metformin can be inferred from the data in Figs. \ref{fig9}(a), \ref{fig9}(b), \ref{fig9}(d) and \ref{fig9}(e). The scatter plot of aspect ratio versus particle length for the metformin particles obtained from the inline PVM images at $T_1$ is truncated at a particle length of about $800\mu$m as seen in Fig. \ref{fig9}(a), whereas a similar estimate from the starting material using the offline Morphologi instrument extends to a particle length of about $3000\mu$m as seen in Fig. \ref{fig3}(e). This is because these long rod-like particles mostly touch the the PVM image frame (see section 2 of the supplementary information for sample images) and hence are rejected by the image processing algorithm. This situation also shows up in the estimated volume based PSD from the PVM images at $T_1$ whose (noisy) peak occurs at a particle length significantly less than that of the corresponding estimate from the starting material using the offline Morphologi instrument seen in Fig. \ref{fig9}(d). Similarly, the peak of the volume based PSD occurs at a CE diameter much less than that of the corresponding estimate of the starting material in Fig. \ref{fig9}(e).

However, the volume based PSDs estimated from the PVM images at $T_3$ (Fig. \ref{fig9}(a)) show better agreement with similar estimates of the milled product using the offline Morphologi instrument as seen in Figs. \ref{fig9}(d) and \ref{fig9}(e). 
However, the volume based PSD of the milled product suggests a higher proportion of fines as seen in Figs. \ref{fig9}(d) and \ref{fig9}(e). This is similar to the case of paracetamol in Figs. \ref{fig8}(d) and \ref{fig8}(e). Similar to that case, the resolution limit of the PVM images and the fact that no reliable estimate of the PSD from the PVM images could be made at $T_4$ (where further breakage may have occurred) could have been responsible for the discrepancy. There could have also been further breakage of the particles during filtration and drying before offline analysis of the milled product.

The estimated aspect ratio PDF of metformin particles both of the starting material and milled product obtained with the offline Morphologi instrument have their peaks close to 0.8. Similarly, the aspect ratio PDF estimated from the inline PVM images has a peak close to 0.8 as seen in Fig. \ref{fig9}(c). This clearly shows that the long rod-like particles of metformin do not dominate the aspect ratio estimated from images. This is in contrast to the mean aspect ratio $r=0.3$ estimated from the CLD at $T_1$ which is dominated by the long rod-like particles.

\section{Conclusions}
\label{sec5}

We have employed wet milling processes of slurries of different crystalline materials to assess the strengths and limitations of two different inline particle monitoring modalities namely CLD and PVM imaging. The materials were carefully chosen as they produce crystals of different morphologies and mechanical strengths.
The work has shown the relative sensitivities of the two modalities to changes in particle size and shape due to the wet milling. The effect of bubbles produced during the process on the two modalities has also been studied.

When the particles are sufficiently large and within the resolution of the PVM instrument camera, and not too large so as to fit into the image frame, the inline imaging method gives PSD estimates which are closer to offline estimates of similar materials. This is especially true for systems composed of long needle-like particles mixed with shorter or more rounded particles. This is because the CLD method needs to find a compromise PSD which fits the measured CLD at an appropriate aspect ratio. 
In systems like these, the aspect ratio distribution estimated by the imaging method tends to be dominated by the smaller particles, while the mean aspect ratio estimated by the CLD method tends to be dominated by the long needle-like or rod-like particles.

However, the inline imaging method is limited to particles of sizes above about $30\mu$m, whereas the CLD method can go to smaller sizes. In addition, the PSD estimated by the inline imaging method becomes less representative as the sizes of the particles approach the size of the image frame. The accuracy of the PSD estimate is also affected by the proportion of objects that are captured outside the focal plane of the camera. Similarly, the PSD estimated by the CLD method becomes less representative as the length of needle-like or rod-like particles approach the diameter of the circular trajectory of the FBRM laser spot making the chord length probability estimate less accurate.

Hence, in systems composed of a mixture of needle-like or rod-like particles and more rounded particles of various sizes, a combination of both the CLD and inline imaging methods (probably in a multi-objective approach) should give more robust estimates of the PSD and the aspect ratio. This will be particularly important to real-time monitoring and control of crystallisation processes where various process conditions could lead to the production of particles of various sizes and shapes and effects such as bubbles which could be very challenging to capture by a single sensor method.

\section*{Data management}
All images captured with the PVM sensor as well as the data from the offline Morphologi instrument and CLD data from the FBRM sensor have been deposited in the publicly accessible repository \href{http://dx.doi.org/10.15129/33e74309-d91c-4ebc-9b2a-18fd33b24876}{University of Strathclyde-Pure}.

\section*{Acknowledgement}

This work was performed within the UK EPSRC funded project \\
 (EP/K014250/1) `Intelligent Decision Support and Control Technologies for Continuous Manufacturing and Crystallisation of Pharmaceuticals and Fine Chemicals' (ICT-CMAC). The authors would like to acknowledge financial support from EPSRC, AstraZeneca and GSK. The authors are also grateful for useful discussions with industrial partners from AstraZeneca, GSK, Mettler-Toledo, Perceptive Engineering and Process Systems Enterprise.

 \newpage
 
 \setcounter{section}{0}

 \setcounter{equation}{0}
 
 \setcounter{figure}{0}

 \setcounter{footnote}{0}
 
 \begin{center}
 \LARGE{\textbf{Supplementary Information}}
 \end{center}
  
\section{Sample images from offline Morphologi instrument}
\label{sec1}

 \begin{figure}[tbh]
\centerline{\includegraphics[width=0.5\textwidth]{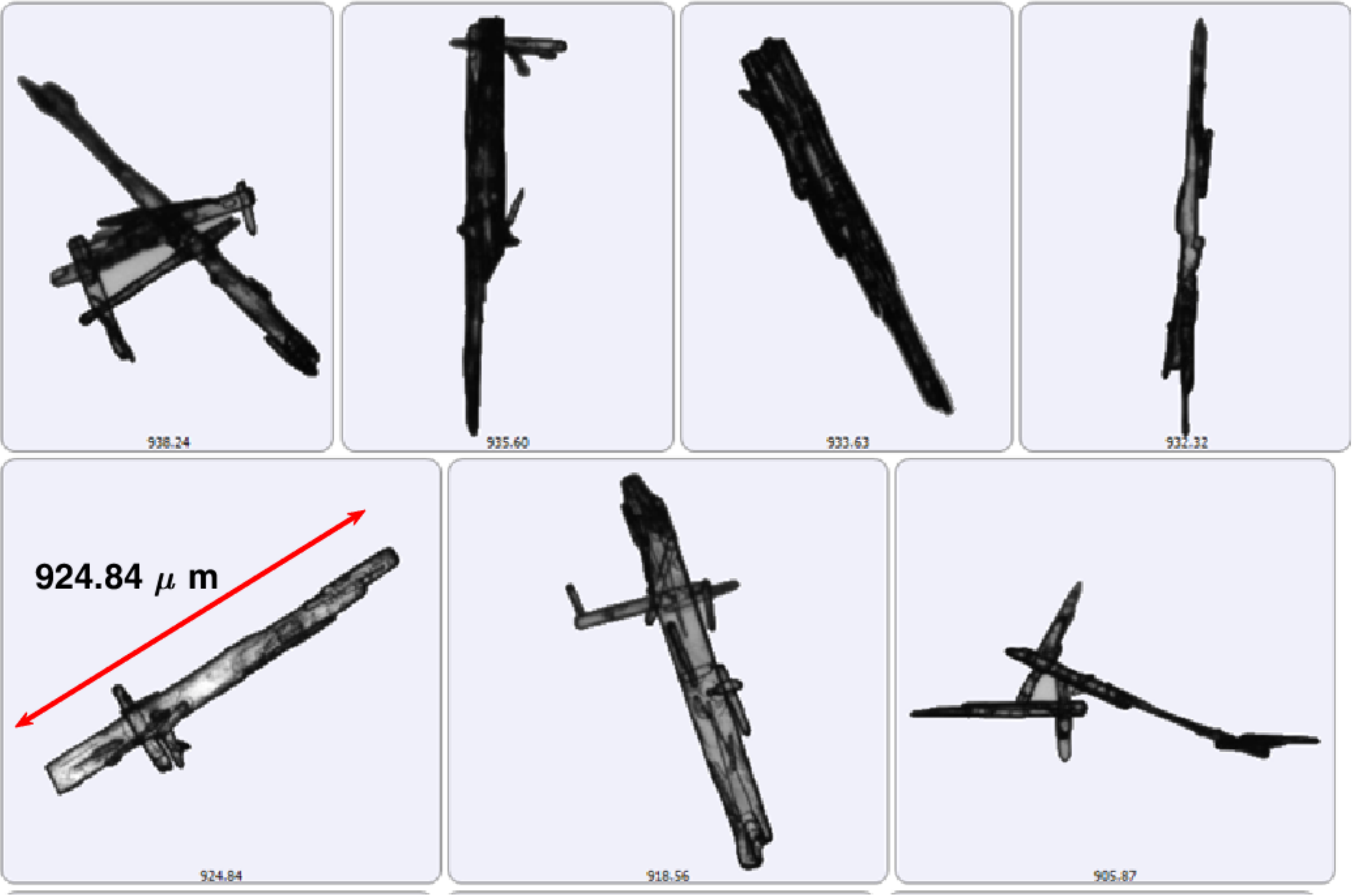}
\includegraphics[width=0.5\textwidth]{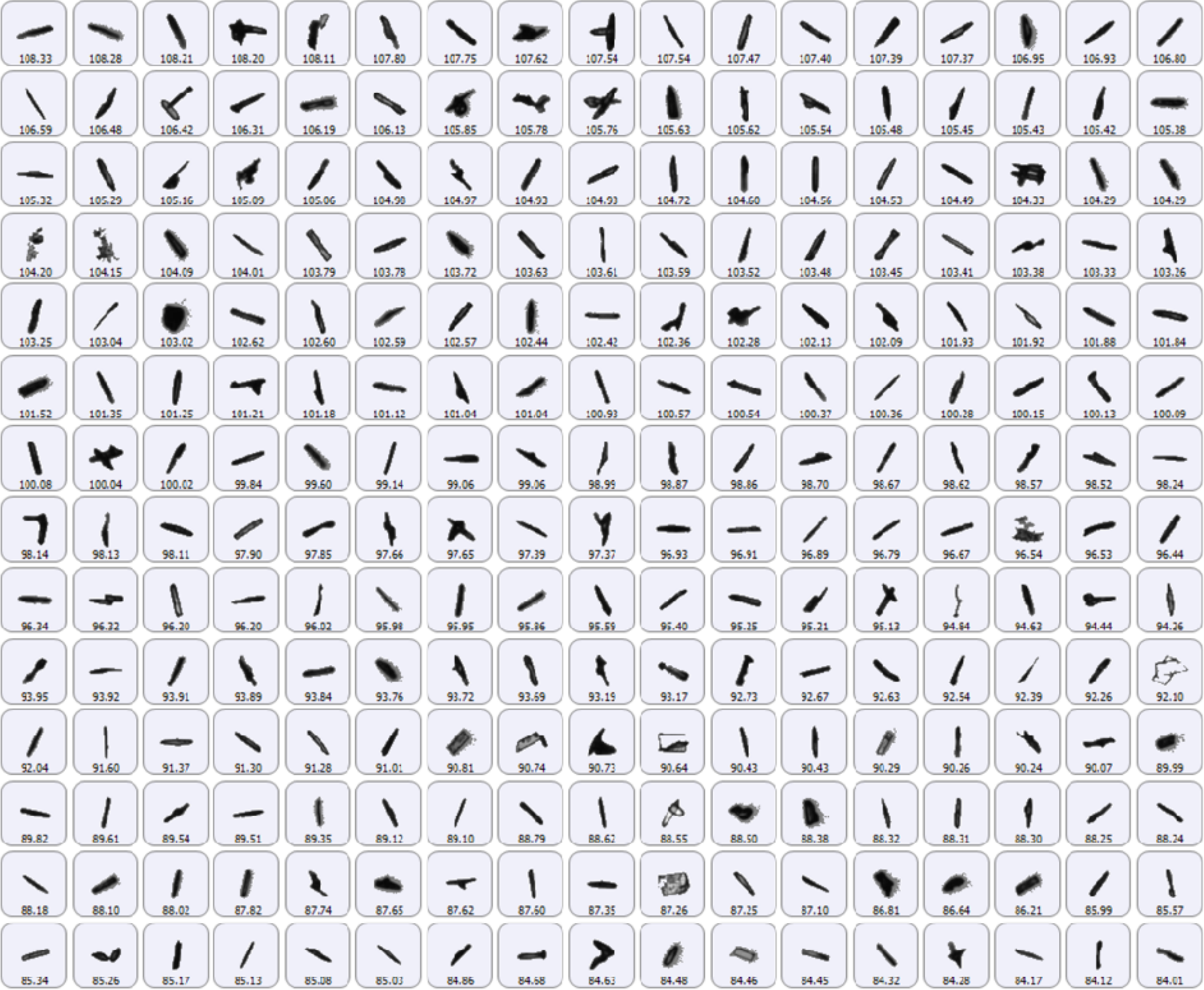}}
\caption{Sample images obtained with the offline Morphologi instrument for the benzoic acid starting material. The object on the left indicated with the double red arrow has a length of $924.84\mu$m. All images in Figs. \ref{figs1} to \ref{figs6} are on the same scale.}
\label{figs1}
 \end{figure}

 \begin{figure}[tbh]
\centerline{\includegraphics[width=0.5\textwidth]{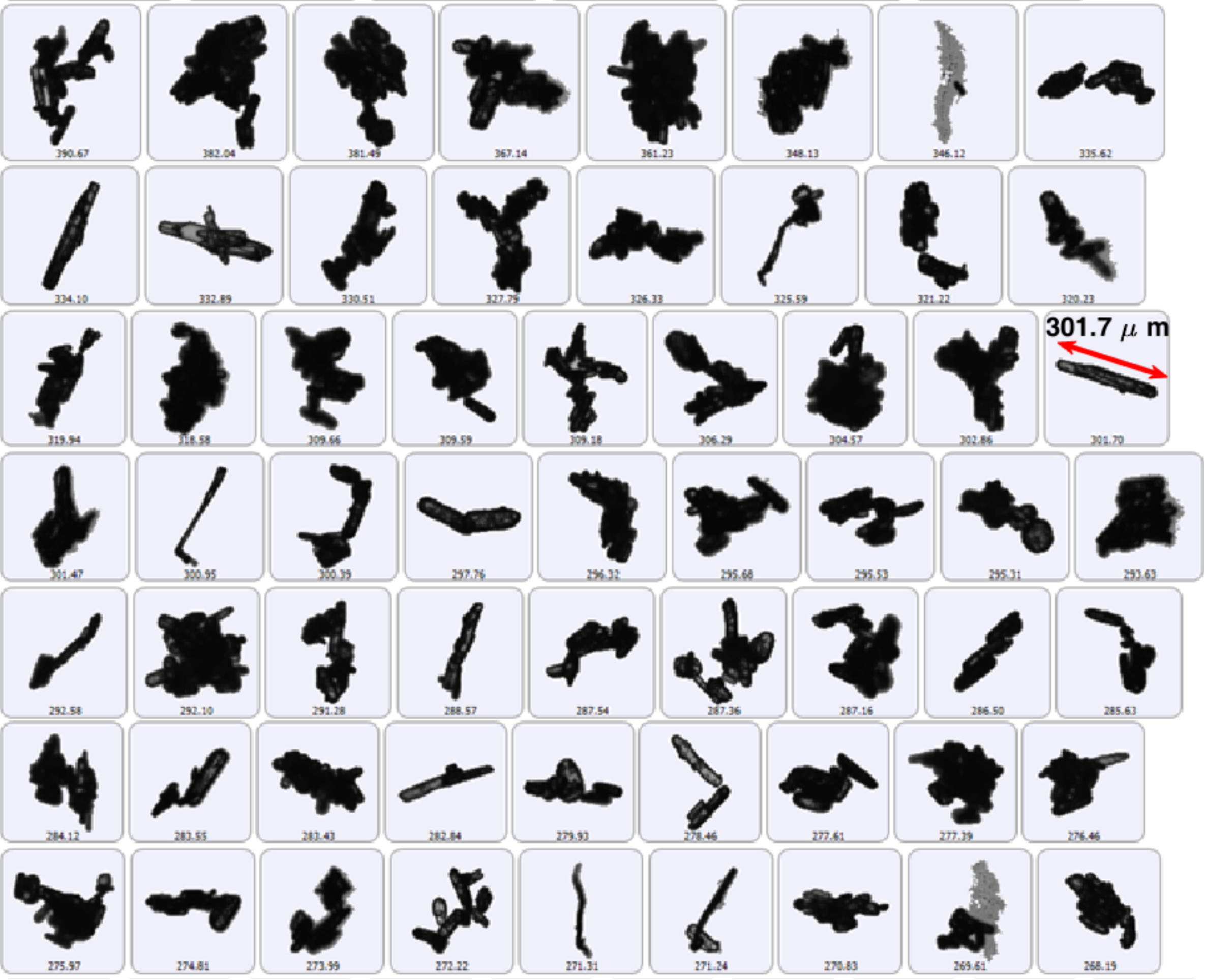}
\includegraphics[width=0.5\textwidth]{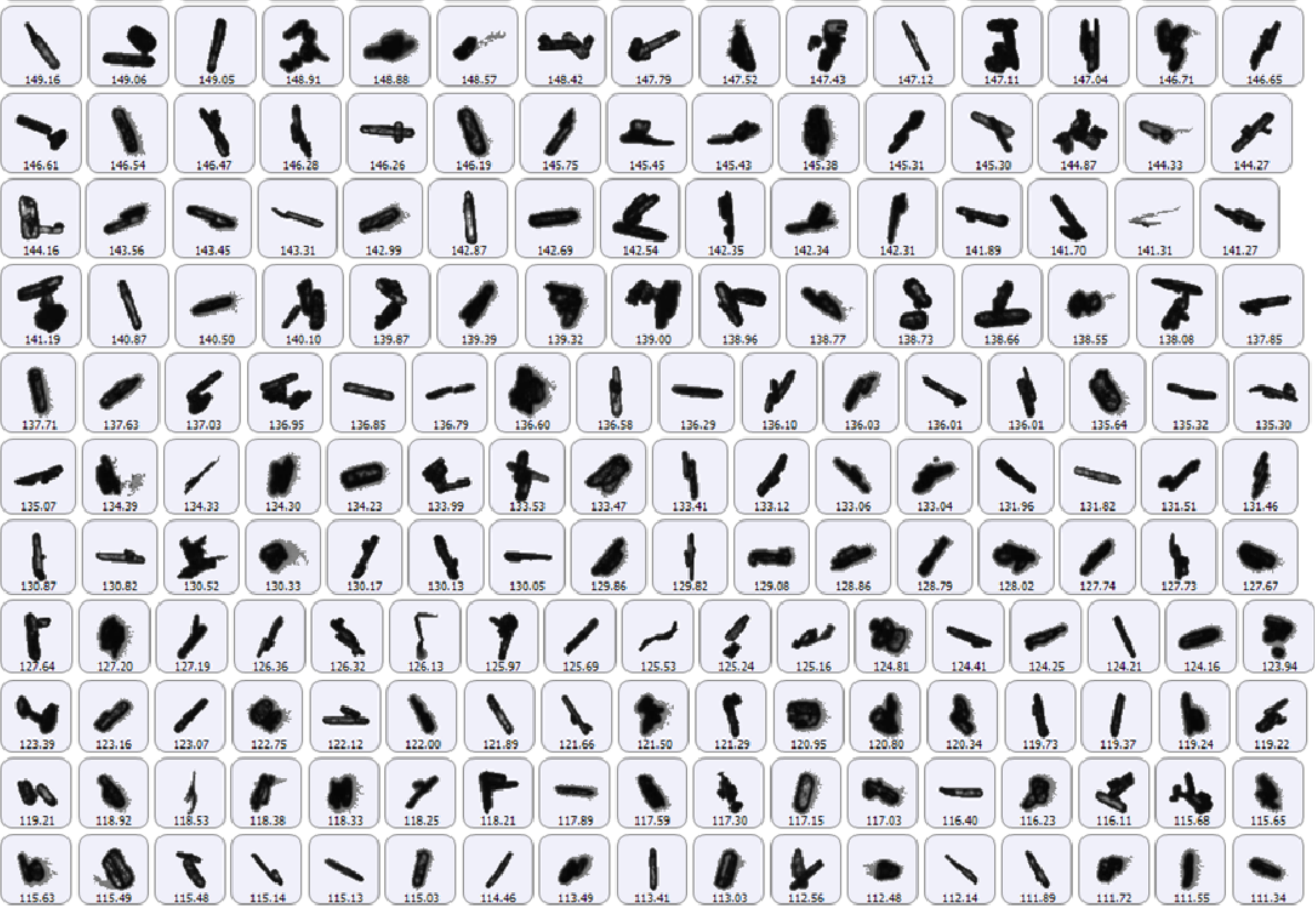}}
\caption{Sample images for the benzoic acid milled product similar to Fig. \ref{figs1}. The object indicated with the red double arrow has a length of $301.7\mu$m.}
\label{figs2}
 \end{figure}

 \begin{figure}[tbh]
\centerline{\includegraphics[width=0.5\textwidth]{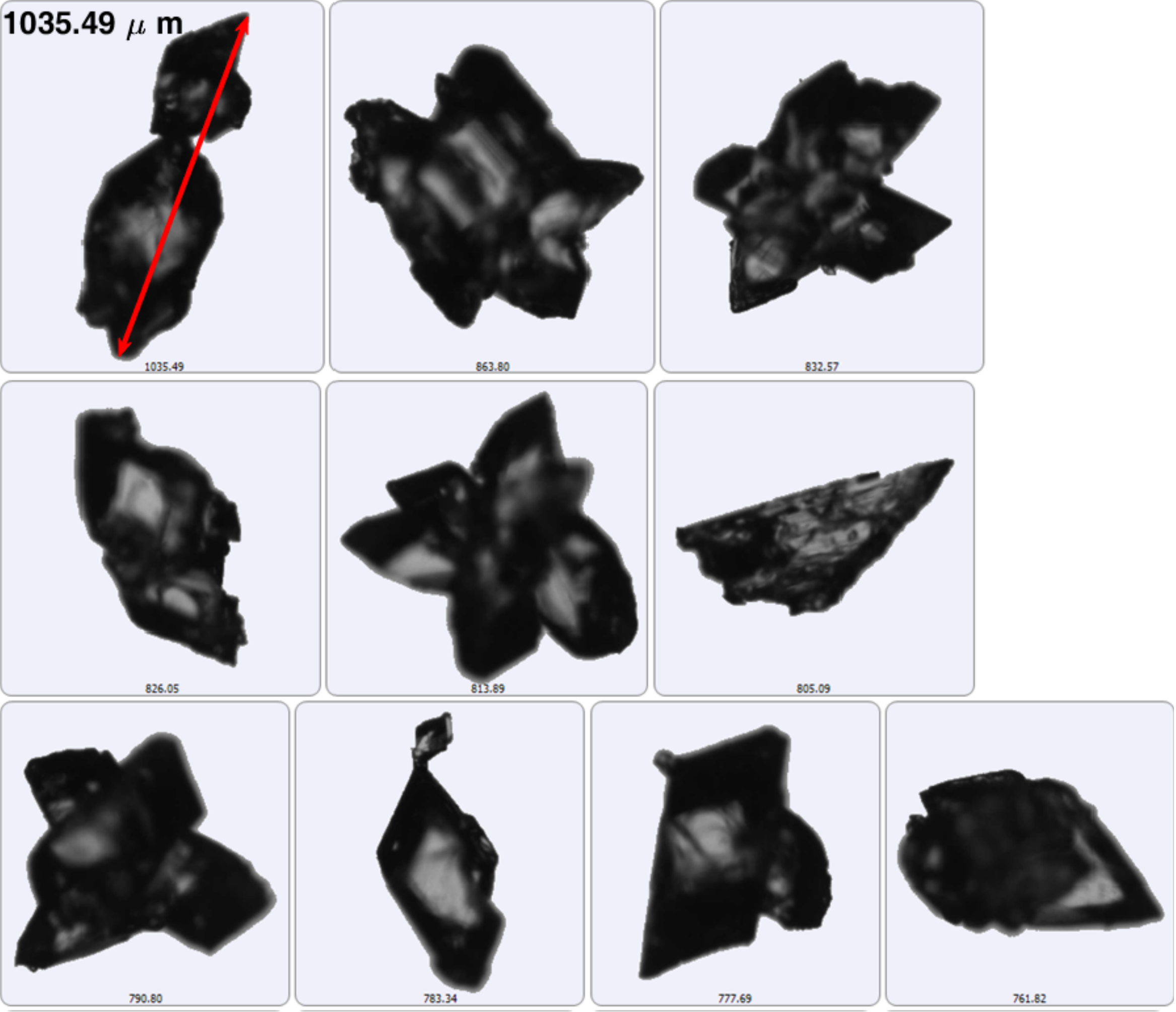}
\includegraphics[width=0.5\textwidth]{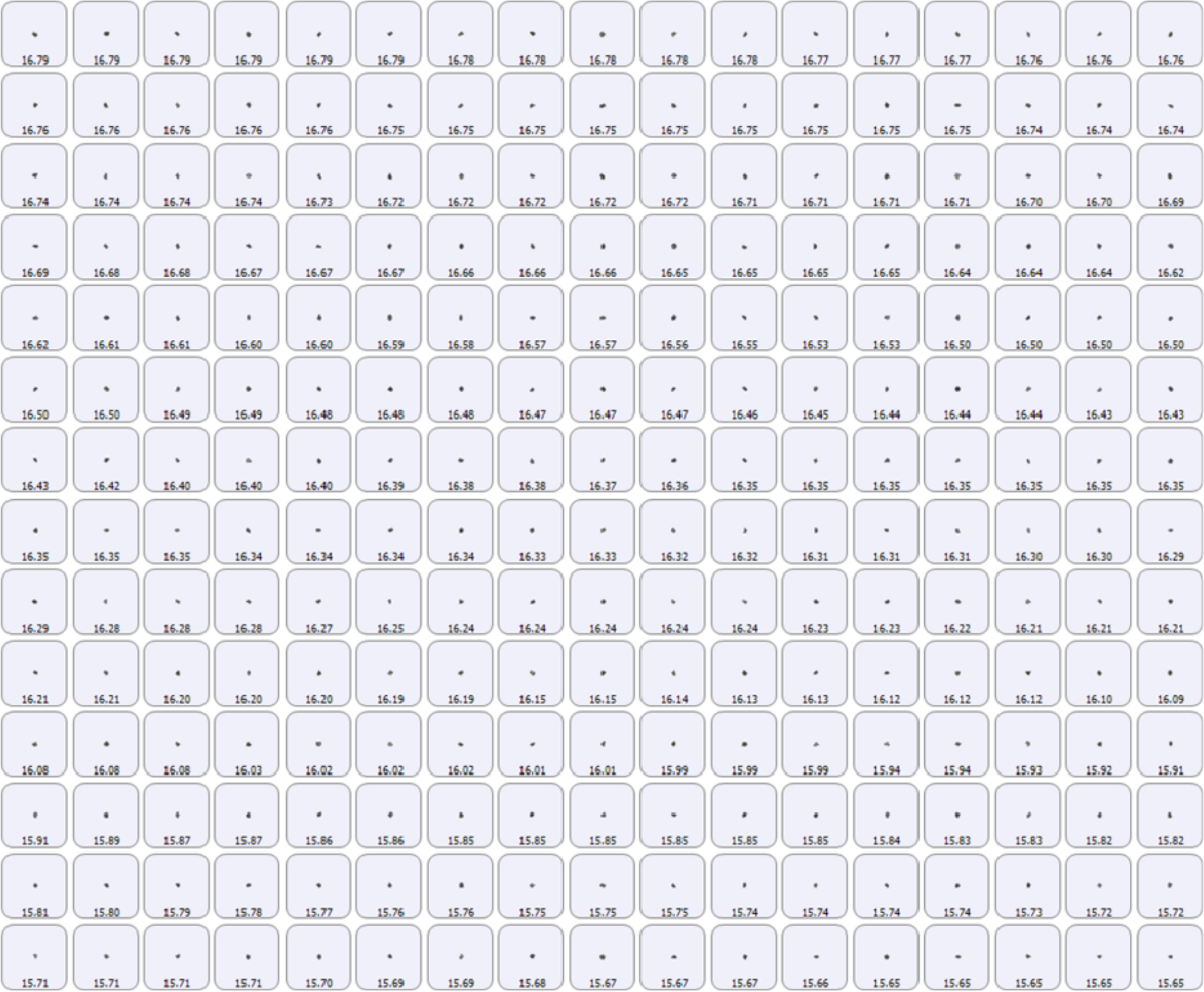}}
\caption{Sample images similar to those in Fig. \ref{figs1} but for the paracetamol starting material. An object with a length of $1035.49\mu$m is indicated with the red arrow.}
\label{figs3}
 \end{figure}

 \begin{figure}[tbh]
\centerline{\includegraphics[width=0.5\textwidth]{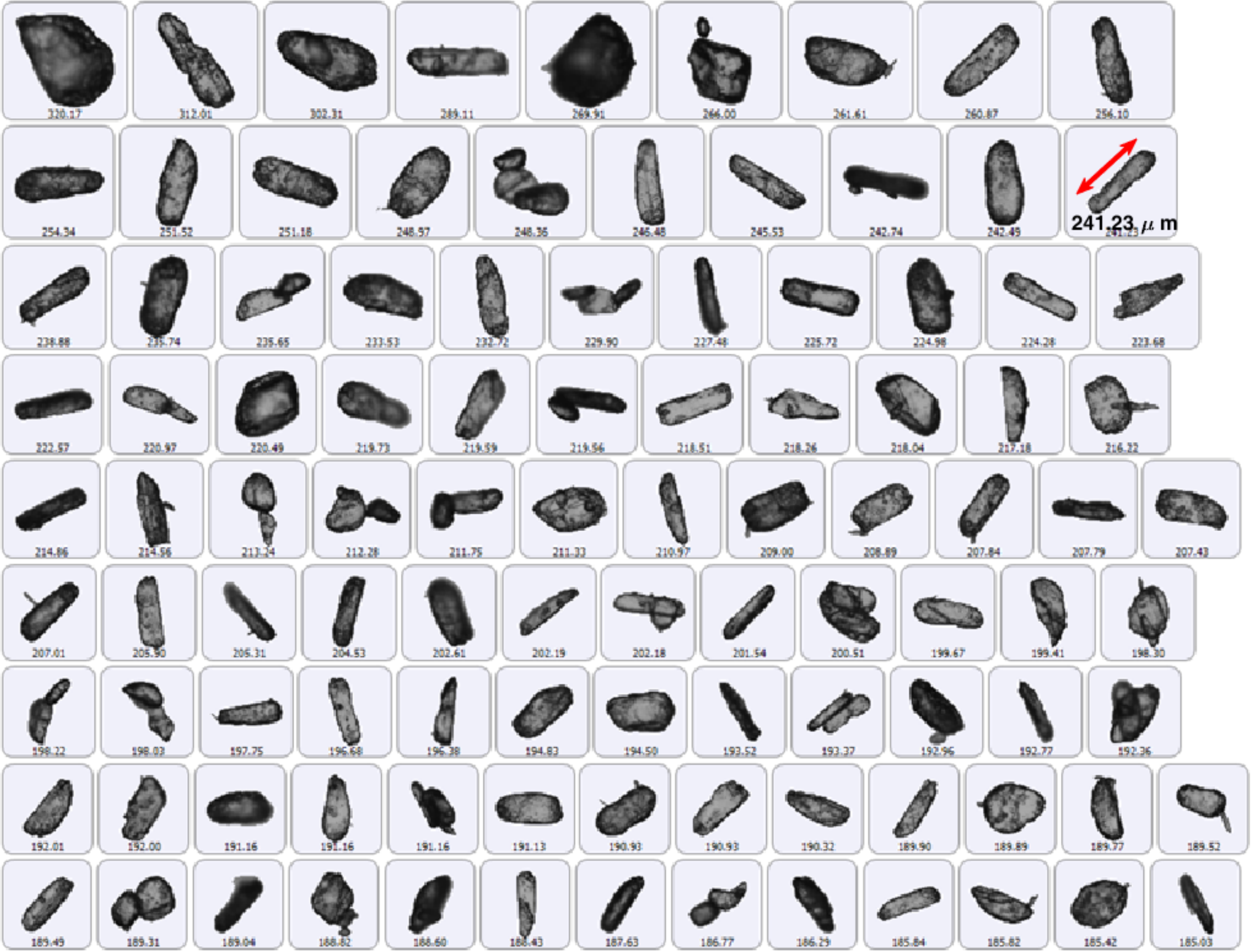}
\includegraphics[width=0.5\textwidth]{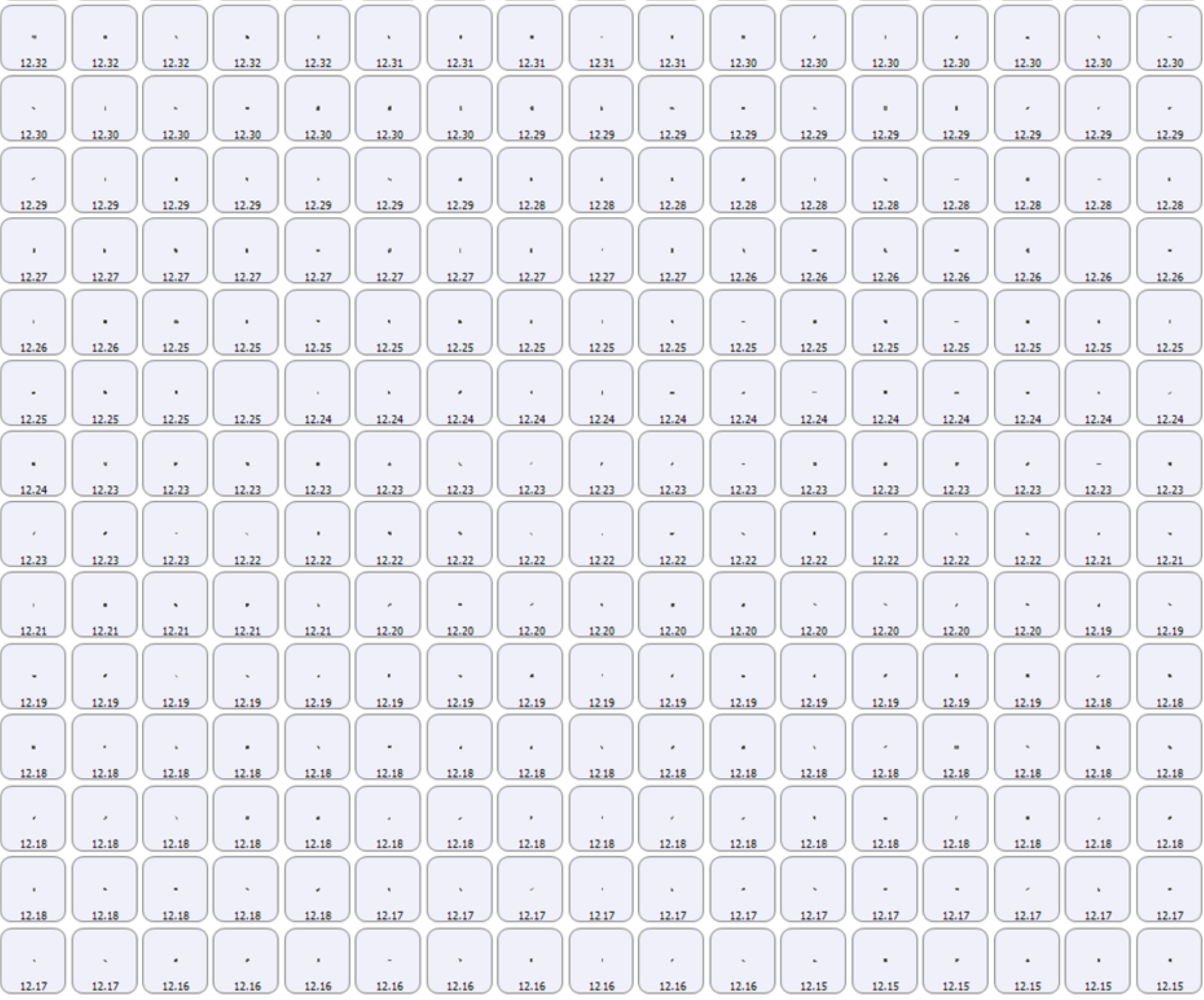}}
\caption{Some images of the milled product for paracetamol similar to Fig. \ref{figs1}. An object of length $241.23\mu$m is indicated with the red arrow.}
\label{figs4}
 \end{figure}

 \begin{figure}[tbh]
\centerline{\includegraphics[width=0.5\textwidth]{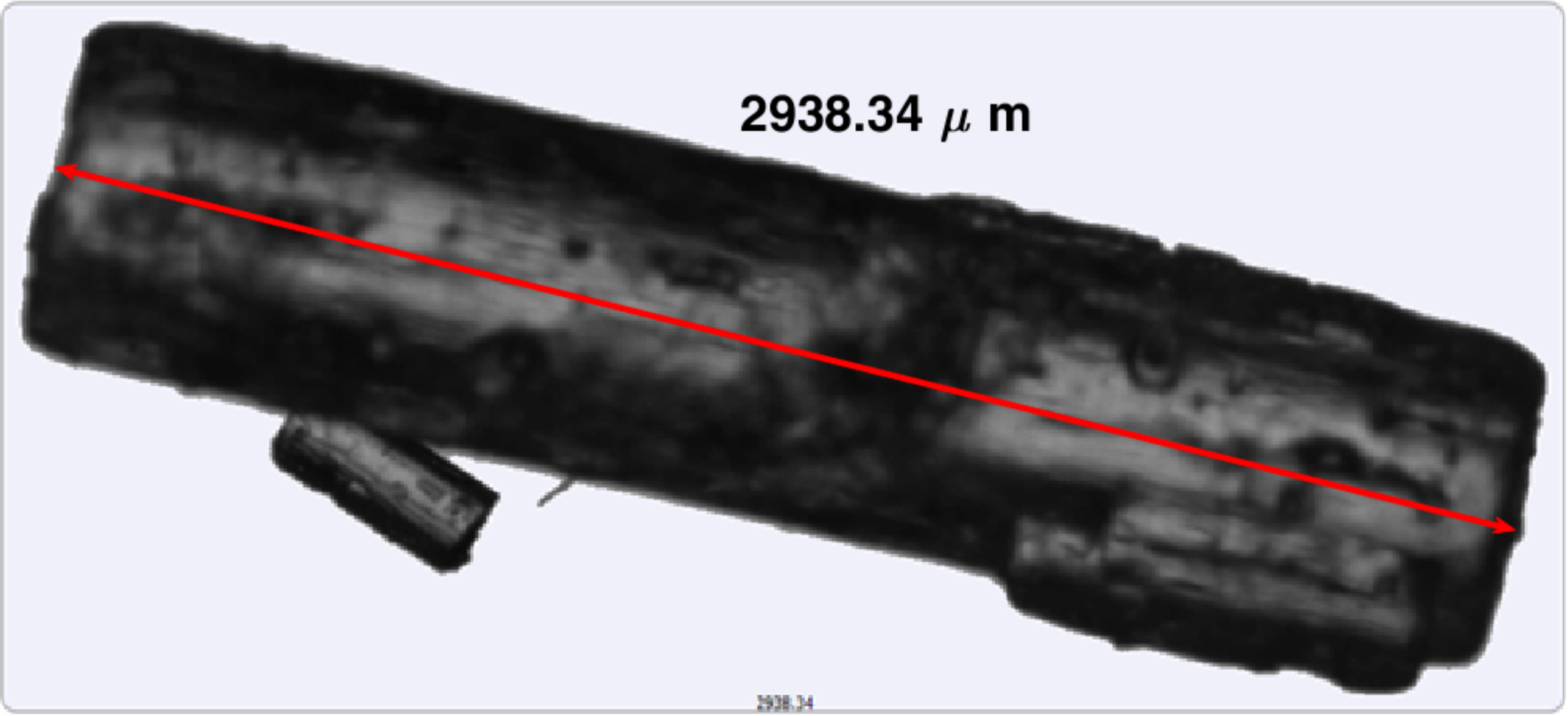}
\includegraphics[width=0.5\textwidth]{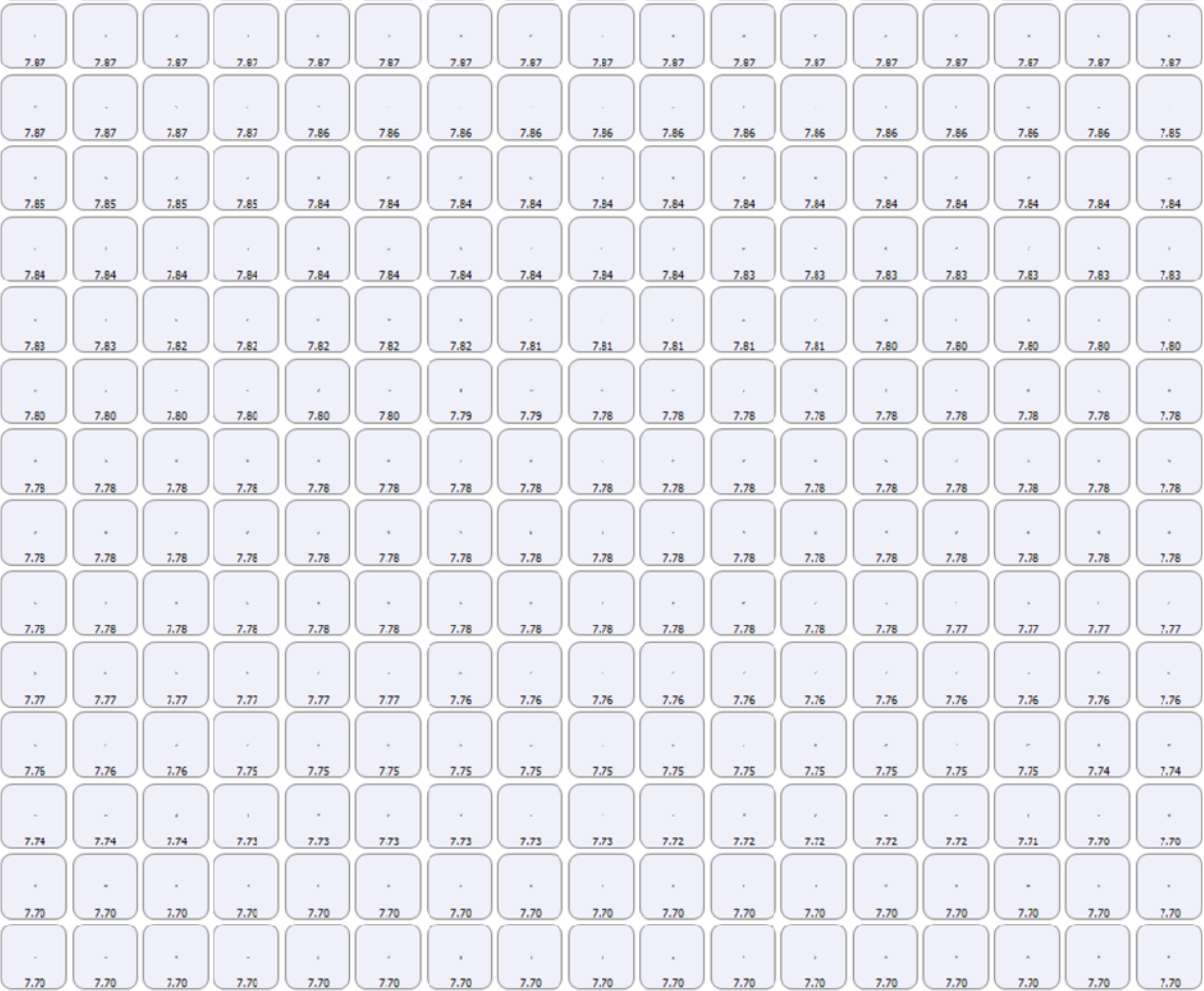}}
\caption{A long rod-like particle and some fines in the starting material of metformin are shown on the left and right. The long rod-like particle has a length of $2938.34\mu$m as indicated.}
\label{figs5}
 \end{figure}

 \begin{figure}[tbh]
\centerline{\includegraphics[width=0.5\textwidth]{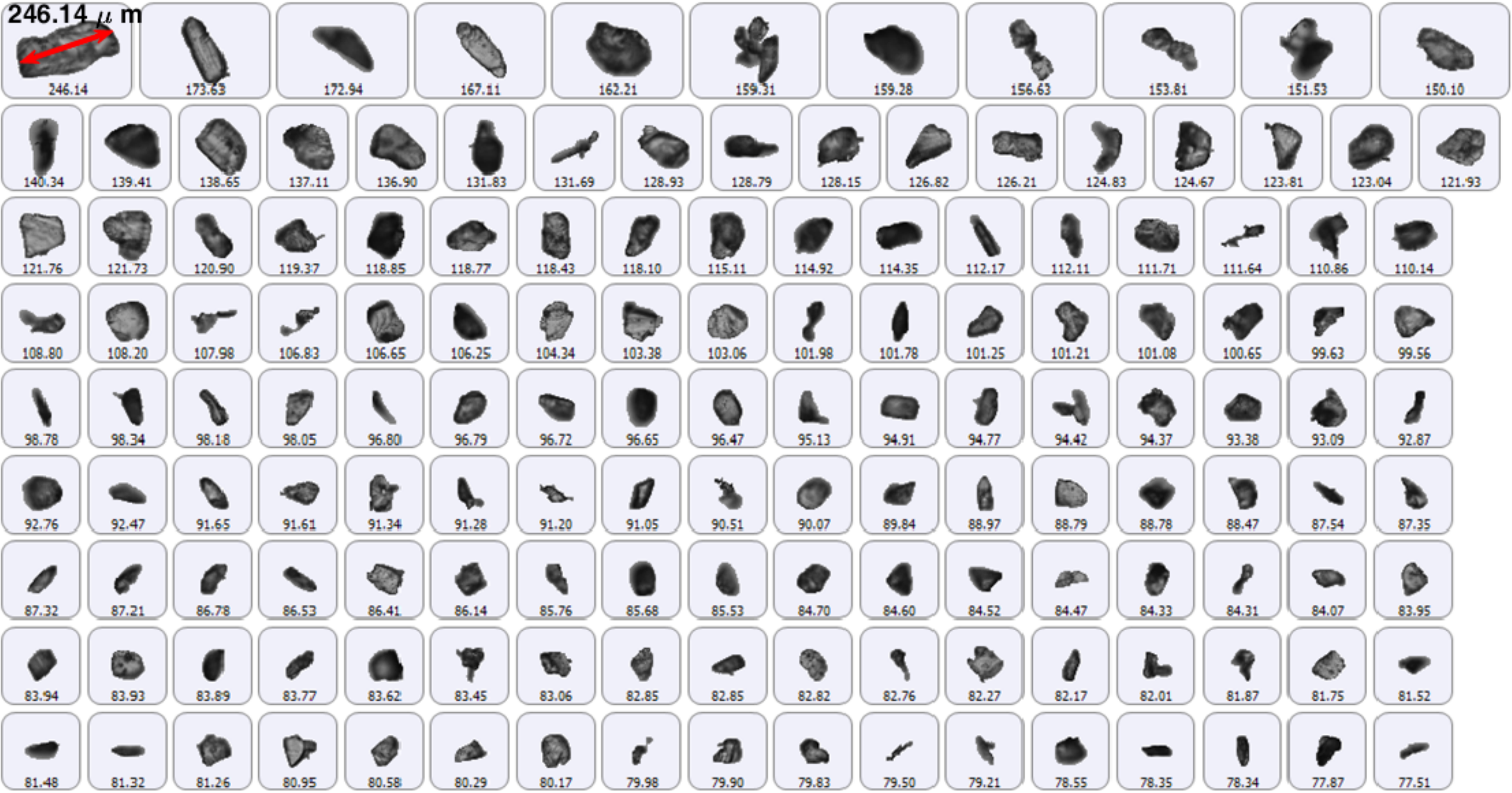}
\includegraphics[width=0.5\textwidth]{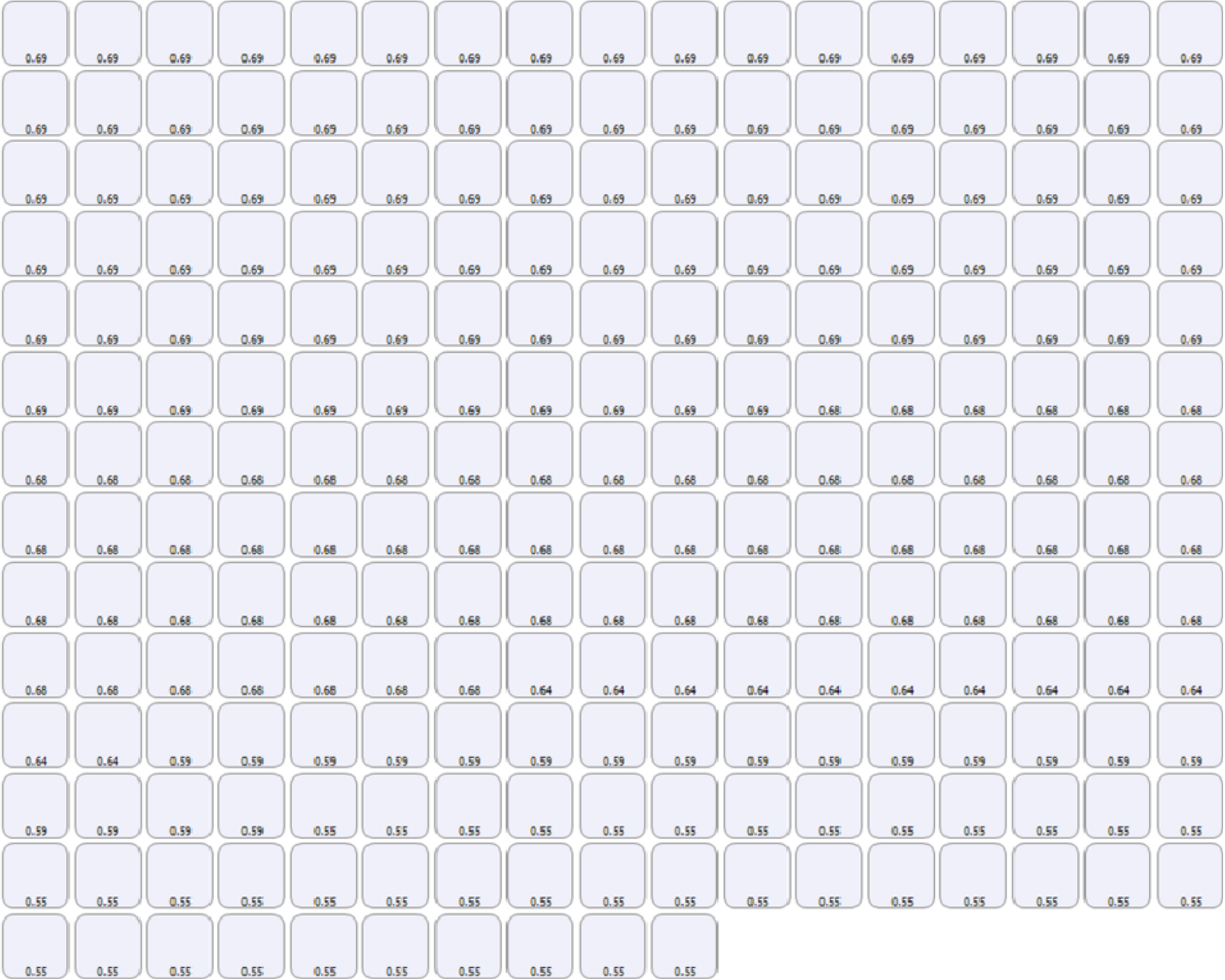}}
\caption{Some of the particles in the milled product (obtained with the offline Morphologi instrument) of metformin are shown in the images. The object indicated with the red double arrow has a length of $246.14\mu$m.}
\label{figs6}
 \end{figure}

Sample images collected for the starting materials and milled products for benzoic acid, paracetamol and metformin are shown in Figs. \ref{figs1} to \ref{figs6}. The images were captured with the offline Morphologi instrument. The images for the starting material for benzoic acid in Fig. \ref{figs1} shows that some of the particles in the starting material in the benzoic acid sample had some degree of agglomeration. The particle sizes also covered a wide range as seen on the left and right of Fig. \ref{figs1}. The degree of agglomeration of the benzoic acid particles had increased by the time the milled product was produced as seen in Fig. \ref{figs2}. 

Similar to benzoic acid, the starting material for paracetamol had particles in different states of agglomeration and covering a wide range of sizes as seen on the left and right of Fig. \ref{figs3}. However, the state of agglomeration of the milled product of paracetamol (Fig. \ref{figs4}) is significantly less than that of benzoic acid in Fig. \ref{figs2}.

An example of the long rod-like particles of the metformin starting material can be seen on the left of Fig. \ref{figs5}. The longest rods had lengths close to $3000\mu$m as seen on the left of Fig. \ref{figs5}. However, the metformin starting material contained a significant amount of fines with particles of lengths as small as around $8\mu$m as seen on the right of Fig. \ref{figs5}. The rods of the metformin starting material were mostly separate with no significant agglomeration. The milled product of the metformin sample contained small particles of lengths $\lesssim 300\mu$m as seen on the left of Fig. \ref{figs6}. However, this milled product also contained a significant amount of fines of lengths $\lesssim 1\mu$m as seen on the right of Fig. \ref{figs6}. The amount of these fines must have been quite huge during the stage $T_4$ of the wet milling process for metformin that the larger particles hardly showed up on the inline PVM images. Hence the inline PVM images were mostly blank during the stage $T_4$ of the wet milling process for metformin, so that the process was terminated at this stage.

\clearpage
\section{Sample images from inline PVM}
\label{sec2}

\begin{figure}[tbh]
\centerline{\includegraphics[width=0.7\textwidth]{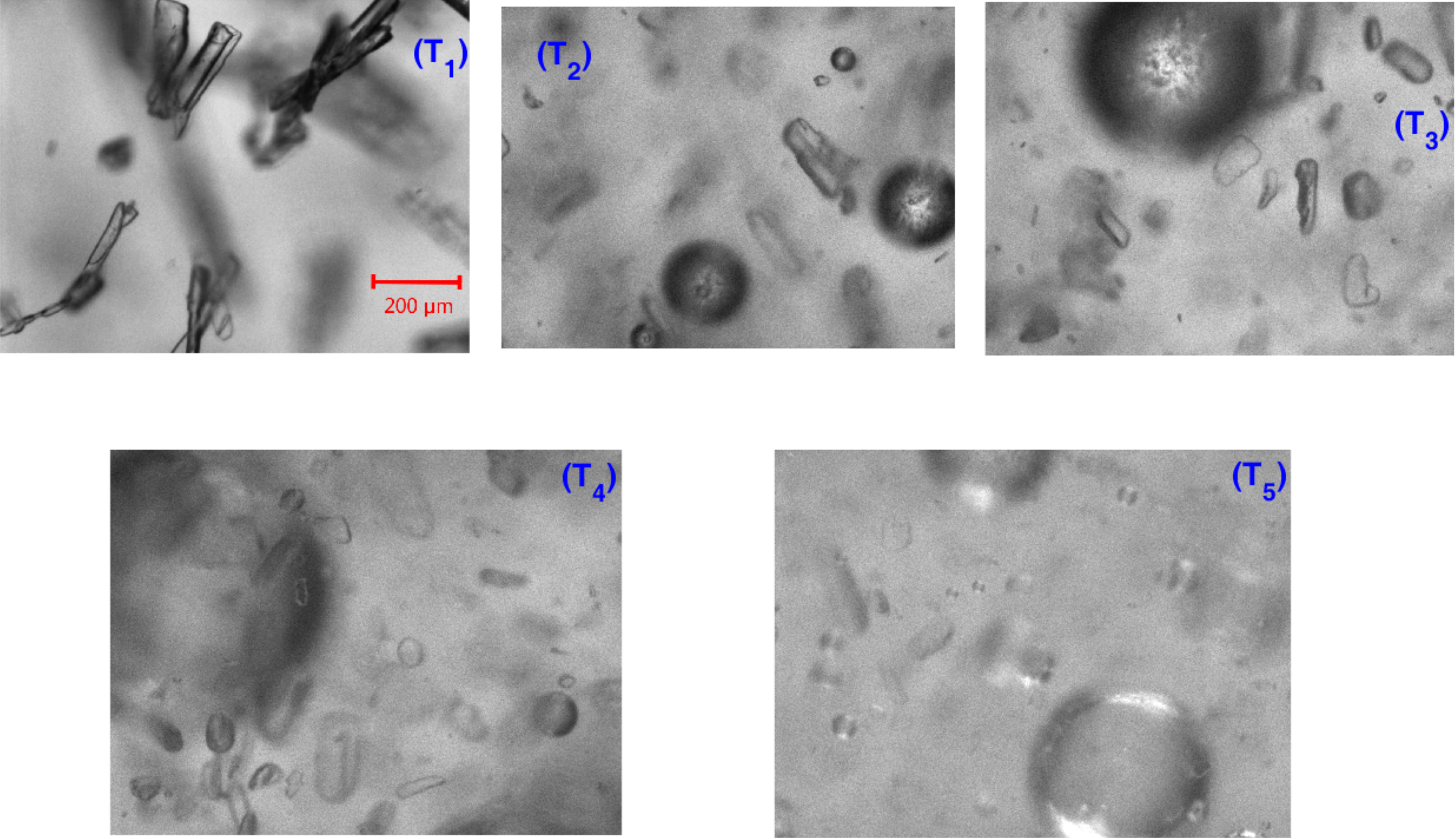}}
\caption{Some sample images collected using the inline PVM sensor during the wet milling of benzoic acid. The images were collected during the stages $T_1$ to $T_5$ of the process as indicated in the images.}
\label{figs7}
 \end{figure}

Some sample images collected with the inline PVM sensor during the wet milling of benzoic acid are shown in Fig. \ref{figs7}. The images in Fig. \ref{figs7}($T_1$) to \ref{figs7}($T_5$) were collected during the stages $T_1$ to $T_5$ respectively of the wet milling process for benzoic acid. The image in Fig. \ref{figs7}($T_1$) shows that some of the particles were agglomerated during stage $T_1$ of the process. The bubbles produced during the wet milling of benzoic acid can be seen in Figs. \ref{figs7}($T_2$) to \ref{figs7}($T_5$). However, the images show breakage of particles.

\begin{figure}[tbh]
\centerline{\includegraphics[width=0.7\textwidth]{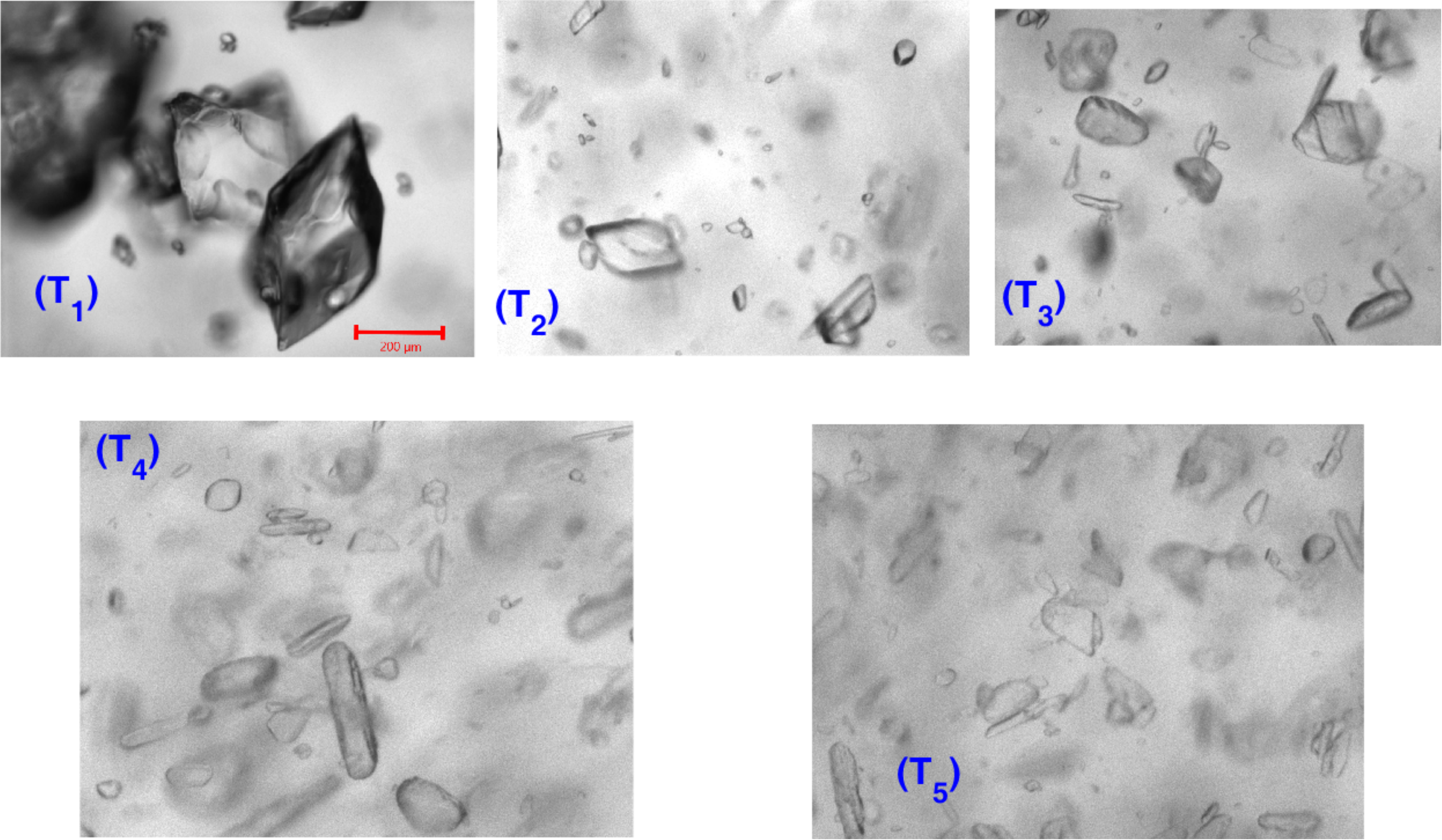}}
\caption{Sample images for paracetamol similar to Fig. \ref{figs7}.}
\label{figs8}
 \end{figure}

Similar to that of benzoic acid are sample images collected during the wet milling of paracetamol in Fig. \ref{figs8}. The images clearly show breakage of particles moving from the stage $T_1$ to $T_5$ during the wet milling process.

\begin{figure}[tbh]
\centerline{\includegraphics[width=0.7\textwidth]{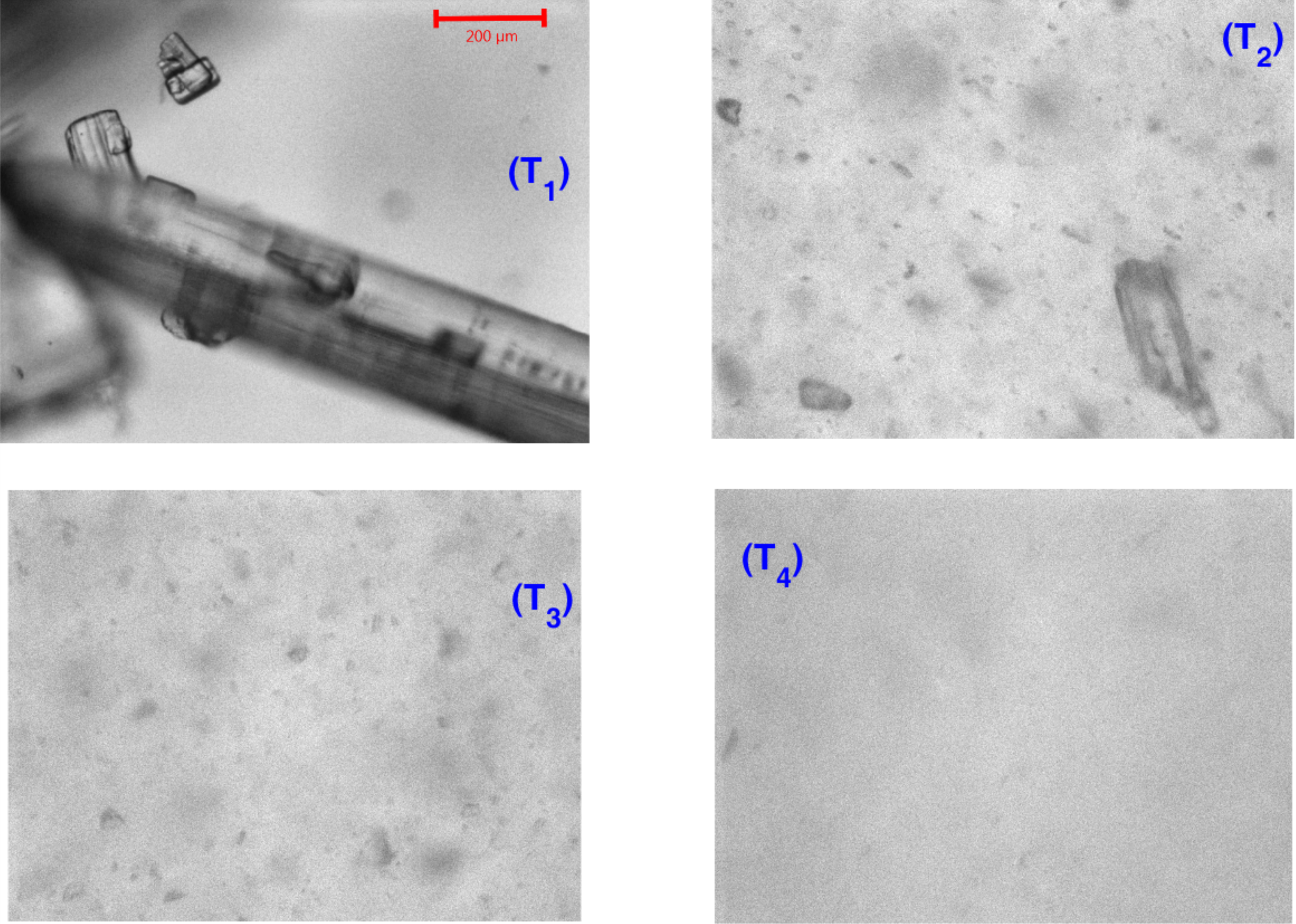}}
\caption{Sample images for metformin similar to Fig. \ref{figs7}. The wet milling process for metformin was terminated at stage $T_4$ as the inline PVM images had almost become blank at the latter part of this stage.}
\label{figs9}
 \end{figure}

In the case of metformin, the particles had broken down considerably moving from stage $T_1$ (Fig. \ref{figs9}($T_1$)) to stage $T_2$ (\ref{figs9}($T_2$)). The long rod-like particle which mostly crossed the image boundaries (Fig. \ref{figs9}($T_1$)) were rejected from the image analysis but contributed to the CLD collected by the FBRM sensor. The led to an estimation of an aspect ratio of 0.3 at stage $T_1$ for the CLD inversion of metformin in the main text. The particles had broken down so much at stage $T_4$ (Fig. \ref{figs9}($T_4$)) that there was no significant contrast between the objects and image background. Hence the wet milling process for metformin was terminated at stage $T_4$. Although, the FBRM sensor collected CLD data at stage $T_4$, the images collected at this stage could not be analysed due to poor contrast. Hence only estimates in the stages $T_1$ to $T_3$ from the images of metformin were reported in the main text. However, estimates from CLD data up to $T_4$ were reported.

\clearpage

\end{document}